\documentclass[times]{elsarticle}
\usepackage{moreverb} 
\usepackage[colorlinks,bookmarksopen,bookmarksnumbered,citecolor=red,urlcolor=red,pdfauthor=author]{hyperref}
\usepackage{amsfonts,amsmath,amssymb}
\usepackage{psfrag,pstricks,pst-node}

\usepackage[section]{placeins}
\usepackage{subfigure} 
\usepackage{mathrsfs}
\usepackage{graphicx}
\usepackage[ruled,vlined]{algorithm2e}
\usepackage{fancyhdr}  
\usepackage{psfrag} 
\usepackage[export]{adjustbox} 
\usepackage{enumerate} 
\usepackage{comment}
\usepackage{lmodern}
\usepackage{media9}
\usepackage{hyperref}
\usepackage{algorithmic}
\usepackage{bm}
\usepackage{booktabs}
\usepackage{threeparttable}
\usepackage{listings}
\usepackage[margin=2.5cm]{geometry}
\graphicspath{{./figs/}{./model structure/}}
\DeclareGraphicsExtensions{.pdf,.eps}

\usepackage{pgfplots}
\pgfplotsset{compat=1.13}

\begin{document}
\makeatletter
\def\ps@pprintTitle{}
\makeatother
 
%\linenumbers
\begin{frontmatter}
\renewcommand{\thefootnote}{\fnsymbol{footnotemark}}

\fancypagestyle{plain}{
\fancyhf{} 
\fancyhead[RO,RE]{\thepage} 
}

\title{An Intrinsic Integrity-Driven Rating Model for a Sustainable Reputation System}
%\title{An Intrinsic Integrity-Driven Rating Model for Sustainable Reputation Oracles}
    \author[lab1]{H. Wen}
    \author[lab2]{T. Huang}
    \author[lab3]{D. Xiao\corref{cor1}} 
    \cortext[cor1]{Corresponding author}
    \ead{xiaodunhui@tongji.edu.cn}
    \address[lab1]{Department of Economics, University of Bath, Bath BA2 7AY, United Kingdom} 
    \address[lab2]{Faculty of Business and Law, University of Roehampton, London, SW15 5SL, United Kingdom} 
    \address[lab3]{School of Mathematical Sciences, Tongji University, Shanghai, China, 200092}  
    
\begin{abstract}

In the era of digital markets, the challenge for consumers is discerning quality amidst information asymmetry. While traditional markets use brand mechanisms to address this issue, transferring such systems to internet-based P2P markets—where misleading practices like fake ratings are rampant—remains challenging. Current internet platforms strive to counter this through verification algorithms, but these efforts find themselves in a continuous tug-of-war with counterfeit actions.

Exploiting the transparency, immutability, and traceability of blockchain technology, this paper introduces a robust reputation voting system grounded in it. Unlike existing blockchain-based reputation systems, our model harnesses an intrinsically economically incentivized approach to bolster agent integrity. We optimize this model to mirror real-world user behavior, preserving the reputation system's foundational sustainability. Through Monte-Carlo simulations, using both uniform and power-law distributions enabled by an innovative inverse transform method, we traverse a broad parameter landscape, replicating real-world complexity. The findings underscore the promise of a sustainable, transparent, and formidable reputation mechanism. Given its structure, our framework can potentially function as a universal, sustainable oracle for offchain-onchain bridging, aiding entities in perpetually cultivating their reputation. Future integration with technologies like Ring Signature and Zero Knowledge Proof could amplify the system's privacy facets, rendering it particularly influential in the ever-evolving digital domain.
\end{abstract}

\end{frontmatter}

\section{Introduction}\label{Introduction}
\vspace{-2pt}

Information asymmetry \cite{Morelli1999} introduces challenges for consumers trying to select products or services that best fit their needs. In traditional retail markets, consumers gravitate towards brands they trust based on personal experiences, the experiences of others, or advertisements \cite{Erdem2004}. Each brand embodies a certain value that can be monetized in open markets. Offering products or services that fall short of their targeted consumers' expectations can diminish this brand value. These brand mechanisms \cite{Baek2010} serve to benefit both consumers and producers in retail and even broader traditional market sectors \cite{Bloom2008}.

With the increased convenience and diversity they provide, online products and services are becoming the preferred choice for many consumers. A notable trend in this domain is the rise of third-party P2P offerings on internet platforms \cite{Zhen2022}. This rapidly growing sector includes a wide array of services. Modern internet platforms such as Amazon, eBay, Taobao, Anyvan, ClickMechanic, and Mybuilder have transformed the way third-party service providers engage with consumers. Essentially, these platforms enable vendors to showcase their products or services, allowing customers to easily browse, compare, and purchase.

Central to these operations is the role of customer reviews. A high number of positive reviews can significantly boost the attractiveness of a vendor, influencing other potential customers' purchasing decisions. Conversely, negative reviews can deter future sales. This significant influence has, however, paved the way for manipulative tactics. The prevalence of deceptive practices, such as fake ratings and click farming, is notably higher online compared to traditional markets \cite{Wu2020, Sekar2023, Harrison-Walker2023}. For instance, some vendors might engage in 'review farming,' where they either purchase their products to leave positive reviews or hire individuals to write misleadingly favorable reviews for them. On the darker side, there might be instances of vendors leaving malicious negative reviews for competitors to tarnish their reputation.

Hence, the establishment of reliable brand mechanisms presents substantial challenges. Given this scenario, verifying authentic consumer actions and crafting efficient reputation mechanisms become critical. While online platforms have the capability to employ algorithms to filter genuine from deceptive actions, thus allowing third-party producers to construct a legitimate reputation, the efficacy of such methods remains questionable. A persistent tension exists between the deceptive maneuvers of some third-party producers and the countermeasures introduced by platform verification algorithms. This dynamic essentially forms a cyclical battle. Present solutions might demand trade-offs, potentially compromising user privacy or other desirable attributes \cite{Shukla2023}. While machine learning offers a promising avenue \cite{Vergne2020}, its implementation requires either an initial training phase with assured genuine user data or a substantial development period before it becomes truly effective.

The advent of blockchain technology, pioneered by Satoshi Nakamoto \cite{Nakamoto2008}, and augmented by Nick Szabo's concept of the smart contract \cite{Cheng2018}—brought to fruition by Vitalik Buterin through Ethereum in 2014 \cite{Ethereum}—has paved the way for innovative on-chain reputation systems.\footnote{In our model, the term "reputation system" is used generically to denote the alternative solution we propose. However, in practical application scenarios, it might also be referred to as a "platform." The two terms can be used interchangeably.} Oracles play a pivotal role in bridging the gap between the real-world and blockchain technology, as they provide the crucial link by which off-chain data can be accessed and utilized by on-chain smart contracts. This function is imperative given that blockchains natively cannot access information outside their network. In 2017, Chainlink introduced its oracle mechanism \cite{Ellis2017}, significantly amplifying the capabilities of decentralized systems. Its renowned use-case for price feeds laid the foundation for the blossoming of Decentralized Finance \cite{Caldarelli2021, Chung2023}. Essentially, oracles act as third-party intermediaries, relaying external data to smart contracts. Chainlink's decentralized oracle network assures the authenticity and integrity of data relayed into the blockchain. This is achieved by sourcing data from a plethora of independent oracles and consolidating their responses, mitigating risks associated with singular data points and enhancing the reliability and security of smart contracts dependent on external data.

Yet, the functioning of these oracles primarily relies on incentive mechanisms, most notably vesting incentives—taking, for instance, the native asset "Link" transitioning from non-circulation to circulation. Over a period, these incentives might have the potential to dilute the value for all participants. As the pool of these rewards nears depletion, apprehensions about the system's enduring viability come to the fore. Successive enhancements, like Chainlink 2.0 \cite{Breidenbach2021}, along with other contemporary solutions, have endeavored to surmount this challenge. Nonetheless, their effectiveness is a subject of continued discussion. While these refined models and strategies demonstrate potential, their ability to maintain prolonged and unwavering effectiveness remains under examination \cite{Wall2021}. Hence, forging a truly sustainable oracle infrastructure becomes essential to fortifying the link between the real-world and blockchain technology.

Recent studies have illuminated various dimensions of blockchain technology, notably the blockchain-based reputation systems \cite{Almasoud2020, Fernandes2023}. While these endeavors are commendable, some of them appear more theoretical and might encounter challenges when catering to all dynamic agent\footnote{In our model, we employ the term "agent" to represent entities involved in the system. Conversely, in real-world application scenarios, we refer to them as "users". Throughout this paper, the term "agent" is used in the context of the model while "user" is employed in the context of real-world applications. Both terms can be considered interchangeable in many scenarios.} interactions. Others have opted for centralized selection of trusted nodes, potentially introducing vulnerabilities associated with single points of failure \cite{Yaqoob2022, Catalini2020, Yakubov2018, Abdullah2017}. Mechanisms aiming to emulate real-world settings sometimes lean on intricate processes for identifying agent types and monitoring their on-chain activities. These processes occasionally overshadow more direct incentive mechanisms that have demonstrated spontaneous efficacy \cite{Atzori2016}. Some proposed economic incentive structures seem to lean on increasing blockchain asset "numbers", which might be susceptible to dilution over time \cite{Metelski2022}, instead of furnishing concrete advantages. As mechanism designs grow in complexity, they might inadvertently elevate intervention costs or introduce unpredictabilities. Relying heavily on uncertainties akin to those proposed by chaos theory \cite{Murphy2010}—inspired by the renowned "Butterfly Effect" \cite{Lorenz1963}—might tread on delicate ground.

While there appear to be promising solutions, such as "Truth" \cite{Qi2022}—a blockchain-aided reputation system aimed at obviating the need to trust e-commerce platforms and ensuring feedback authenticity and anonymity—the system is not without its potential shortcomings. "Truth" introduces a hybrid model wherein the centralized e-commerce platform remains in charge of traditional trading services, but the blockchain oversees the authentication of feedback. Although "Truth" presents an innovative approach to the age-old problem of authenticating e-commerce feedback, certain aspects warrant deeper scrutiny. For example, the paper does not adequately address the motivations that would drive consumers to engage in the feedback process. Understanding these incentives is vital for ensuring widespread adoption and sincere feedback. Furthermore, the system seems to overlook potential malicious behaviors by sellers, such as posting negative feedback against competitors or hiring fictitious consumers to generate positive reviews. While transactions may take place through the prescribed channels, there is a potential for compensating these "hired" consumers via alternative methods, thereby circumventing the system's safeguards. For "Truth" to fully realize its potential and to demonstrate its effectiveness in real-world settings, these concerns need to be addressed in future versions or research.

Given these challenges, there's merit in exploring a more streamlined, basic model that inherently incentivizes agent behavior integrity and fosters sustainability \cite{Zheng2020, Li2023, Han2023, Xu2021, Liu2022}. Initial evaluations of this model could employ Monte-Carlo simulations across vast parameter domains, capturing real-world intricacies. It's essential to note that this rudimentary model need not capture every intricate detail of real-world scenarios. A basic reputation system, even if it offers only a singular credibility rating, can be invaluable in myriad applications. This value is realized when the reputation information it delivers is accurate and trustworthy, allowing users to rely on it rather than turning to other platforms that struggle to filter out low-cost fake ratings, as previously discussed.

Building upon a well-defined basic model foundation allows for iteration and enhancement. By integrating corrective components to the foundational equations, we can preserve core incentives while refining the model towards an everlasting reputation system to mirror real-world scenarios more accurately. Ensuring this model's competitive advantage over other platforms or blockchain solutions amplifies its resilience and sustainability. This strategy not only highlights the model's practical utility but, as adoption grows, leverages the network effect. Widespread integration of our system fortifies commercial barriers against replication, making it increasingly arduous for potential competitors to emulate our unique model and its benefits.

While the applications of the everlasting reputation system continue to expand across various domains and demographics, this system can function as a sustainable oracle network with demonstrated reliability. The integration of Ring Signatures \cite{Fujisaki2007, Zhang2002, Wang2008, Chow2005, Melchor2011} and Zero Knowledge Proofs (ZKPs) \cite{Fiege1987, DeSantis1992, Kilian1992} can further enhance the privacy dimensions of such oracle mechanisms, especially in situations where privacy is paramount. Ring Signatures allow for digital signature from a group member without disclosing the identity of the specific signer. ZKPs represent an advanced cryptographic technique enabling agents to substantiate certain claims (like their assets or states) without disclosing the specifics.

%\subsection{Structure}
The structure of this paper is organized as follows: Section \ref{Relevant Blockchain Technologies} introduces the relevant blockchain technologies. Following this, the basic model is presented in Section \ref{Basic Model}, which lays the foundation for understanding the incentive mechanisms that promote self-motivated integrity among agents. We then refine this basic model in Section \ref{Advanced Model} to mirror real-world scenarios more closely, ensuring that the incentives for integrity-driven ratings are preserved. After building a robust theoretical base, Section \ref{sec:Monte_Carlo} uses Monte Carlo simulations to validate our propositions, followed by a detailed analysis and discussion of the results. Our findings from the simulations are summarized in Section \ref{sec:Conclusions}, where we underscore that our reputation system functions effectively as a universal sustainable oracle. Finally, potential directions for future research are proposed in Section \ref{sec:future_investigations}.

\section{Relevant Blockchain Technologies}\label{Relevant Blockchain Technologies}

Introduced in 2009 as the foundation of Bitcoin, blockchain technology is a decentralized, immutable distributed ledger, facilitating the transparent and verifiable recording of transactions. Each transaction is appended to a structure known as a 'block', which in turn links to the previous block, thereby forming a chain.

The advantages of blockchain are manifested in its transparency, security, decentralization, and immutability. Despite its transformative potential across sectors—such as supply chain management, finance, healthcare, and voting systems—blockchain grapples with various challenges. These encompass high energy consumption, scalability, transaction speed, and regulatory issues. A notable challenge, particularly salient for integrating the real-world with blockchain, is the lack of sustainable oracles. Oracles serve as critical infrastructure, bridging the gap between on-chain and off-chain environments, yet their sustainability and reliability remain areas of concern. Nevertheless, the optimism surrounding blockchain's ability to revolutionize diverse industries persists.

In this section, we will delve into the workings of \textbf{Oracles as the Connections between Blockchain and the Real World} and \textbf{Blockchain in Voting Systems}, discuss the \textbf{Scalability Requirements} our reputation system necessitates, and enumerate relevant \textbf{Blockchain Technologies for Potential Integration}.

\subsection{Oracles as the Connections between Blockchain and the Real World}

Oracles, in the realm of blockchain, function as pivotal bridges connecting the real world with on-chain smart contracts. These systems essentially imbue blockchain platforms with a capability they intrinsically lack: accessing and processing data from external sources. When a real-world event or condition is met, oracles convey this information to smart contracts, triggering predefined operations. In such a setting, the credibility, timeliness, and accuracy of the data become indispensable, as the execution of the contract relies on this external information.

Yet, ensuring the sustained and trustworthy operation of oracles presents challenges, a significant one being the incentive mechanism. To encourage data providers to supply accurate and timely information, many oracle systems deploy tokens as rewards. However, this approach often entails the transition of native assets, like "Link", from non-circulation to circulation. Over time, this can potentially dilute the value of the tokens for all holders. As the reservoir of these incentives approaches depletion, questions about the long-term sustainability of such a system become pressing. How can one maintain the motivation of data providers without compromising the system's overall stability and value?

This paper aims to address the above conundrum by introducing our novel blockchain reputation system. We posit that this system can function as a sustainable solution for oracle operations. Rather than leaning heavily on token-based rewards, our approach integrates the principles of reputation and trust, ensuring that data providers are motivated by both immediate rewards and the longer-term benefits of maintaining a trustworthy reputation. In the subsequent sections, we will delve into the intricacies of our proposal, illustrating how it offers a harmonious blend of incentives, trustworthiness, and sustainability.

\subsection{Blockchain in Voting Systems}

Blockchain technology, renowned for its decentralized, transparent, and immutable nature, finds a novel application in voting systems, providing a means to enhance the transparency, security, and integrity of elections. This section will delve into the fundamental mechanisms underlying blockchain-based voting and showcase its unique advantages.

\subsubsection{Operational Mechanism of Blockchain Voting}

A blockchain-based voting system operates as follows:

\begin{enumerate}
\item \textbf{Voter Registration:} For traditional blockchain-based voting systems aimed at public elections or polls, participants are typically required to register on the platform. Once verified, voters are provided with a unique cryptographic identity that ensures both anonymity and eligibility. This method mimics the conventional voter registration process but integrates the benefits of blockchain.

However, in the context of blockchain governance, especially on Proof of Stake (PoS) platforms, the registration process may differ. In these scenarios, the act of creating a blockchain wallet and acquiring assets within that wallet often implicitly grants the holder voting rights. The weight of their vote might be determined by the amount of the asset they hold. Thus, there isn't a separate 'registration' as such; instead, participation in the network (by holding assets) inherently provides the ability to vote.

    \item \textbf{Voting Process:} When casting a vote, the voter signs it with their cryptographic key. This vote is then added as a transaction to be confirmed and recorded on the blockchain.
    
    \item \textbf{Vote Counting:} Once the voting period concludes, the votes can be tallied directly from the blockchain. Given the blockchain's immutable nature, the records cannot be changed, ensuring an accurate count.
    
    \item \textbf{Transparency and Verification:} At any point, anyone can verify the votes on the blockchain without revealing the identity of the voter, ensuring a transparent yet confidential system.
\end{enumerate}

\subsubsection{Real-world Implementations}

Many existing blockchain platforms have adopted on-chain governance and voting mechanisms, demonstrating the feasibility and reliability of this approach. For example:

\begin{itemize}
    \item \textbf{Cosmos (ATOM)}: The Cosmos Network utilizes a staking and delegation model wherein ATOM holders can delegate their stakes to validators \cite{cosmosGitHub2019, cosmoswhitepaper2019}. These validators then propose and vote on network changes and upgrades, ensuring decentralized decision-making. Each vote's weight is proportionate to the stake, promoting a sense of accountability among validators.
    
    \item \textbf{Tezos (XTZ)}: Tezos has an on-chain governance system where XTZ holders vote on protocol upgrades \cite{tezoswhitepaper2018, tezoswhitepaper2014}. Instead of contentious hard forks, the platform evolves through community consensus, emphasizing adaptability.
\end{itemize}

To sum up, blockchain's inherent properties—security, transparency, and decentralization—make it an ideal candidate for reshaping voting systems. By ensuring a democratic and trustworthy process, it offers a future-proof solution to traditional voting challenges.

\subsection{Scalability Requirements} \label{Scalability Requirements}

Scalability has perennially been one of the most contentious challenges in the blockchain domain \cite{yang2020, chauhan2018}. It stands as a stumbling block that has led numerous promising projects to falter or even wither away before achieving their potential. The crux of the scalability challenge is the need to handle a vast number of transactions per second (TPS) without compromising on speed, security, or decentralization.

However, our reputation system elegantly sidesteps this vexing issue. The fundamental reason is its modular design: while our system is dedicated solely to ratings or votes, the actions and trades of agents can occur on any external public blockchain. This delineation ensures that the TPS demand for our system is driven exclusively by the frequency at which agents rate or vote. In reality, such activities are infrequent. On average, an individual might cast a rating or a vote once every few days. Even in the most engaged scenarios, a single individual might rate only a few times within a day.

Given this low-frequency engagement, even with participants in the order of millions, the system would only necessitate a TPS at the level of hundreds. Notably, this is well within the capabilities of many existing blockchain platforms \cite{pierro2022, crain2021}. In essence, by confining our system's operations to these particular tasks and allowing other actions to be handled elsewhere, we've effectively decoupled our platform's functionality from the more demanding scalability requirements that have beleaguered so many other projects.

\subsection{Blockchain Technologies for Potential Integration}

\subsubsection{Ring Signatures: Ensuring Anonymity and Authenticity}

Our blockchain-based reputation system can benefit from the technique of Ring Signatures to safeguard the privacy of agents. This ensures that they can express their opinions without concerns of potential retaliation or discrimination, especially when giving negative feedback. 

Ring Signatures, originally introduced for ensuring anonymous communications, plays a pivotal role in ensuring both the authenticity and anonymity of transactions in our reputation system. A ring signature, in essence, is a type of digital signature that can be generated by any member of a group of users, which we'll refer to as a 'ring'. The distinguishing feature of this signature is its ambiguity: while anyone can validate the signature to ascertain that a member of the group generated it, it remains computationally infeasible to determine the exact individual who created the signature.

Within our reputation system, when a user wishes to endorse or rate an entity, they can use ring signatures to validate their authenticity (proving they belong to a group of legitimate users) while keeping their individual identity concealed. This ensures that while ratings and feedbacks are trustworthy and genuine, raters can maintain their privacy.

For instance, if Alice wants to rate a service, she can produce a ring signature using her private key and the public keys of other members in the ring. When others verify this rating, they can be confident that it originated from someone within the group, but they can't pinpoint it specifically to Alice, ensuring her anonymity.

\begin{lstlisting}
interface RingSignature {
    function generateSignature(bytes memory _data, bytes[] memory _publicKeys) public returns (bytes memory);
    function verifySignature(bytes memory _data, bytes memory _signature) public returns (bool);
}
\end{lstlisting}

\subsubsection{Threshold Functions in Zero Knowledge Proofs}\label{Threshold Functions in Zero Knowledge Proofs}

Zero Knowledge Proofs (ZKPs), specifically in our design at a later stage when the computing environment is allow, can be implemented to introduce a threshold functionality. This is a sophisticated feature allowing agents to prove their reputation score exceeds a set threshold without unveiling the exact score, preserving user's privacy. Although valuable, this feature introduces a computational overhead due to the intrinsic complexity of ZKPs. As a starting point, our design opts to unveil interfaces for this threshold functionality without instant deployment, preserving computational efficiency.

\begin{verbatim}
interface ThresholdZKP {
    function verifyProof(bytes memory _proof) public returns (bool);
    // Potential future functions
}
\end{verbatim}

\subsubsection{Commitment Schemes with Rounds}

To accentuate fairness, especially in scenarios where real-time rating or feedback might be infeasible, our system envisages the use of Commitment Schemes \cite{Cai2019, Kim2018} in a round-based manner. In this paradigm, participants first 'commit' to their ratings within a specific round. Following the round's conclusion, these commitments are 'revealed'. This stratagem promotes fairness within each round, eliminating the possibility of preemptive actions based on insights or predictions. If our system doesn't adopt this round-based approach initially, we envision the Commitment Schemes to be woven in at a later stage, bolstering the system's fairness.

\begin{verbatim}
interface CommitmentRounds {
    function commitRating(bytes memory _rating) public;
    function revealRating(bytes memory _commitmentKey) public;
    // Supplementary functions for advanced features
}
\end{verbatim}

In essence, the blend of Ring Signatures, ZKPs, and Commitment Schemes amplifies the robustness and fairness of our blockchain-based reputation system. This blend ensures a future-ready framework that starts lean but is capable of seamlessly adopting advanced features.

\section{Basic Model} \label{Basic Model}
%this section include these subsections: On-chain reputation system; Connecting the real world; Assumptions of the Gains or Losses for Any Agent; Distribution of Agent Actions; Action Analysis in Reputation Systems 

In this section, we embark on a deep exploration of the foundational mechanisms underpinning our reputation-based blockchain. We initiate our discourse with the \textbf{On-chain reputation system}, elucidating how reputation metrics are embedded and maintained directly within the blockchain environment. Progressing further, we tackle the essential task of \textbf{Connecting the real world} to our digital framework, ensuring seamless integration and applicability. Recognizing the importance of clear foundational premises, we detail the \textbf{Assumptions of the Gains or Losses for Any Agent}, shedding light on the anticipated dynamics each participant might experience. This naturally leads us to an examination of the \textbf{Distribution of Agent Actions} within the system, offering insights into the varied interactions and behaviors exhibited by agents. Finally, we delve deep into the \textbf{Action Analysis in Reputation Systems}, providing a granular breakdown of the ripple effects and broader implications of agent activities within our reputation ecosystem.

\subsection{On-chain reputation system} \label{On-chain reputation system}
The idea is to introduce a cross-rating system in the blockchain-based free market. All agent actions will be rated either positively or negatively by others using staked credit points. The incentive for rating lies in potential future profits, as the staked credit points for each comment will earn a profit if it aligns with the general sentiment (positive or negative) of future comments. Conversely, they will incur a loss if future comments contradict the initial sentiment. The more credit points one stakes, the higher the potential profit or loss one stands to gain based on future ratings. Simultaneously, this also amplifies the impact on all previous ratings in the same direction, reinforcing them if they align and diminishing them if they oppose. Here, an agent is defined as a blockchain wallet holder. An agent can choose to stake any amount of its credit points between 0 and the balance in the wallet minus the transaction fee (negligible at Layer 2 of the Ethereum blockchain \cite{Asgaonkar2022, Kudzin2022}). An agent can also distribute the amount across multiple wallets, but this won't provide any advantage over operating as a single entity since weight is determined by credit point amounts, not the number of agents.

\begin{figure}[h!] 
  \centering
  \includegraphics[width=1.0\textwidth]{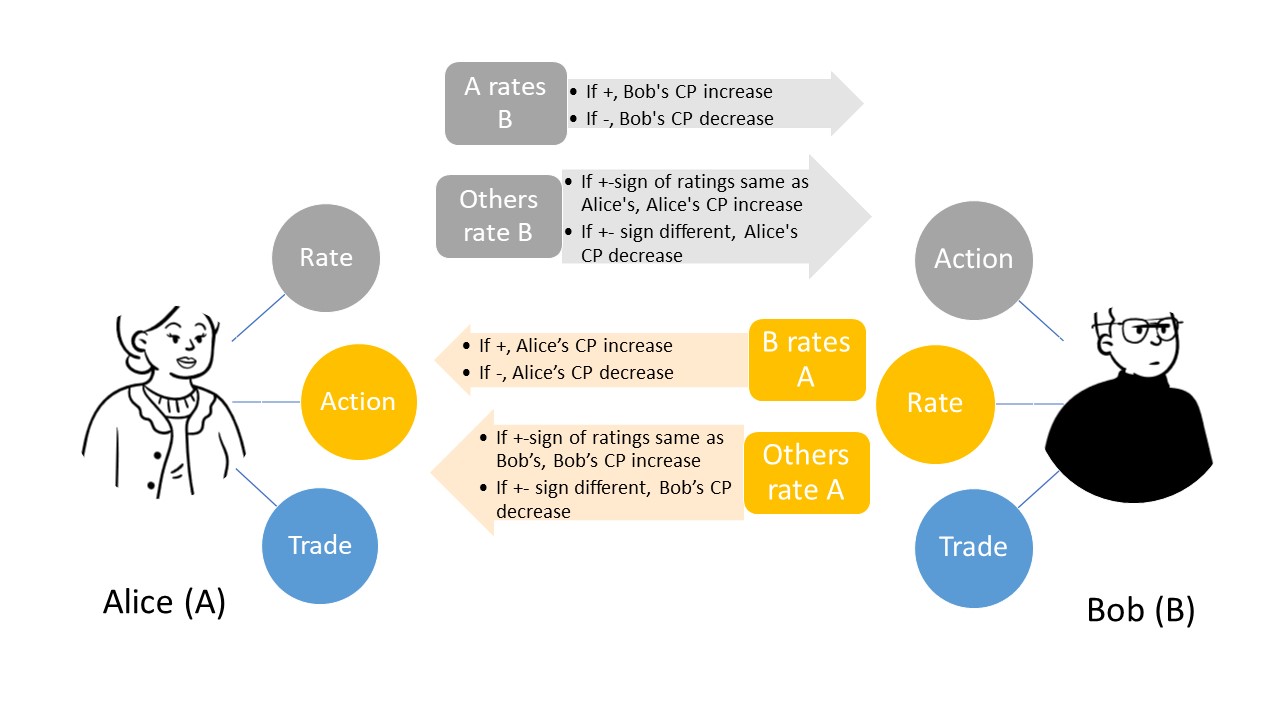}
  \caption{A typical round of agents interactions in the reputation system: Alice and others rate Bob's Action, and in the same round, Bob and others rate Alice's Action}
  \label{fig:A typical round of agents interactions}
\end{figure}

The core proposition is to integrate a cross-rating system into the blockchain-based free market. Every agent's action gets rated—either positively or negatively—by other agents leveraging staked credit points. The incentive for this rating mechanism emerges from potential future gains; the staked points for a comment will appreciate if subsequent comments resonate with its sentiment. In contrast, if future comments diverge, the staked points depreciate. The magnitude of potential gains or losses directly correlates with the number of staked credit points. Moreover, these stakes influence all preceding ratings in the same direction, either amplifying or attenuating them. In this context, an agent is delineated as a blockchain wallet holder. This agent can stake any quantity of its credit points, bounded between zero and the wallet balance after accounting for transaction fees. While agents can distribute their credit points across multiple wallets, this doesn't confer any strategic advantage since the weight stems from the quantity of staked points, not the agent count.

Figure \ref{fig:A typical round of agents interactions} showcases an archetypal interaction round among agents within the reputation system. In any given round or block, agents, be it Alice, Bob, or any other participant, can partake in a mix of actions, ratings, or trades. For instance, if Alice rates Bob's action positively in a round, Bob garners positive returns. On the other hand, a negative rating imposes a loss on Bob. Other agents can similarly rate Bob's action within the round, influencing his credit point balance based on the rating's nature.

When delving deeper into ratings, if an agent's rating of Bob's action aligns with Alice's sentiment, both Alice and that agent benefit. However, conflicting ratings lead to losses for both parties.

The dynamic nature of potential gains and losses, contingent upon future ratings, fosters authentic feedback from agents, mitigating false ratings. Given that a significant portion of these ratings are genuine, the ecosystem offers a tangible advantage over traditional online platforms, which often grapple with counterfeit, low-cost ratings. Over time, this advantage is projected to burgeon, drawing an expanding user base. As the positive reinforcement loop of ratings solidifies, the continuously growing influence of the reputation system will inspire participants to proactively contribute, transforming these ratings into the system's cornerstone. Such inherent positive externalities vouch for the sustainability of this reputation system, setting it apart from numerous other blockchain paradigms that lean heavily on credit point vesting, such as the renowned Chainlink 2.0 \cite{Wall2021}.

\subsection{Connecting the real world} \label{Connecting the real world}
Following the introduction of "soulbound credit points" by Ethereum's co-founder, Vitalik Buterin \cite{Soul2023, Owens2022}, and Elon Musk's vision of integrating blockchain with Twitter accounts \cite{Handagama2022}, a unique opportunity arises to associate each wallet address from the on-chain reputation system with a specific individual or entity in the real world. This association implies that actions undertaken offline could be evaluated, and these evaluations could subsequently result in tangible gains or losses based on evolving collective ratings. As blockchain continues to evolve and integrate with real world including Real World Assets (RWAs) \cite{Zhu2021, Hou2023} via oracle or similar mechanisms, an increasing number of high-value economic activities are likely to find connections to the blockchain.

Given the promising potentials discussed in Subsection \ref{On-chain reputation system}, individuals with a higher risk tolerance might be incentivized to link their real-world actions to their blockchain wallet addresses. On the other hand, participants of the reputation system, essentially the wallet holders, are the ones recognizing the value of this association and are willing to submit their ratings in anticipation of future benefits. This sentiment is particularly pronounced during the initial phases of the project. Although speculators may invest in credit points expecting them to appreciate in value, they don't possess a strong economic motive to submit misleading ratings. As a result, their activities should not compromise the system's functionality.

We'll now transition from these theoretical underpinnings to a practical application, beginning with a rudimentary model. Let's consider the following categories of agents participating in the reputation system:
\bigskip
\\
C1. Agents with Actions\\
C2. Rating Participants\\
C3. Investors
\bigskip

Any given agent can belong to one, two, or all three of these categories simultaneously.

C1 agents, due to potential gains or losses stemming from others' ratings, are more likely to be responsive to rating if they have staked a higher number of credit points. It is logical to assume that agents staking more credit points will prioritize meeting consumer expectations. Since this staking information is transparent, consumers will naturally gravitate towards producers who have staked more credit points, creating an initial incentive for credit points purchase when someone decides to join the reputation system. While producers can also rate consumers, this mechanism naturally promotes responsible action across the board. To preserve individual privacy and provide flexibility in engagement, the model permits C1 agents to choose the amount they would like to stake – anywhere from zero to their full wallet balance (minus transaction fees) – for any given action they want to be rated on. The staking information is transparent on the chain, and therefore consumers may also use this information when choosing providers. They are more likely to prefer those who have staked more credit points due to the reasoning above.

C2 agents similarly determine their staked amounts based on the confidence level of their beliefs when providing ratings. Since these agents are subject to potential gains or losses depending on the conformity of others' ratings to the actions they have rated, they are incentivized to rate in a responsible and genuine manner.

C3 agents, or traders between the credit points and other assets at anywhere external to the reputation system, are individuals who primarily buy or sell credit points without engaging in other actions within the system. Their actions are largely inconsequential to the reputation system's functionity. However, traders could indirectly benefit the system in its initial phase by driving up credit point prices with their purchases.

\subsection{Assumptions of the Gains or Losses for Any Agent} \label{Assumptions of the Gains or Losses for Any Agent}

Consider an agent \(i\) as a general example of any agent. We categorise its involvement in the reputation system into three parts: functioning as a C1, C2, and C3. Considering trader agents categorised as C3 are not directly affecting the credit point-number-based gains or losses of any other specific agents, we now focus on C1 and C2 categories and denote them as "A" for action agents and "R" for agents providing ratings.

For Agent \(i\), let \(S^A_{i_m}\) represent the credit points staked on action \(i_m\) when acting as a producer or consumer, which is always a positive value. Meanwhile, \(S^R_{ik_l}\) represents the credit points staked on rating the action \(k_l\) when acting as a rating provider on the action \(k_l\) of another agent $k$. This value can be positive if the ratings align in direction, and negative if they are opposite.

For action \(i_m\) of agent \(i\), the gains or losses incurred are influenced by any agent \(j\) who provides ratings. This influence is proportional to the amount of credit points staked by both \(i\) and \(j\). This influence is denoted as

	\begin{eqnarray}
		\displaystyle {\Delta CP_{iji_m}^A} &=&\displaystyle {C_{R2A}S^A_{i_m}S^R_{ji_m},} 
		\label{momeqn1CPA} %\\
	\end{eqnarray}
 
where \(\Delta CP_{iji_m}^A\) is the gains of agent \(i\) from its action \(i_m\) as an action agent rated by agent \(j\), \(C_{R2A}\) is the constant coefficient from rating to action.

Concerning the rating \(n_i\) from agent \(i\) on the bahaviour \(k_l\), by \(k\), any agent \(j\) rating the same action by \(k\) (with staked credit points represented by \(S_{j,k_l}^{A}\)) influences the gains or losses. Again, this influence is proportional to the amount of credit points staked by both \(i\) and \(j\). We label this influence as

	\begin{eqnarray}
		\displaystyle {\Delta CP_{ijk_l}^R} &=&\displaystyle {C_{R2R}S^R_{ik_l}S^R_{jk_l},} 
		\label{G_AiR} %\\
	\end{eqnarray}

where \(C_{R2R}\) represents the constant coefficient of a rating's impact on another rating.

In reference to agent \(i\), the cumulative contribution from all other agents can be summarised as:

	\begin{eqnarray}
		\displaystyle {\Delta CP_i} &=&
		\displaystyle {\Sigma_{j,m} C_{R2A}S^A_{i_m}S^R_{j_m} + \Sigma_{j,k,l} C_{R2R}S^R_{ik_l}S^R_{jk_l}},
		\label{momeqn1CPi} %\\
	\end{eqnarray}
where \(\Delta CP_i\) denotes the total gains agent \(i\) accrues from its participation in the reputation system.

We can naturally assign example values to the constant coefficient \(C_{R2A}\) as

	\begin{eqnarray}
		\displaystyle {C_{R2A}} &=&
		\displaystyle {\frac{1}{\Sigma_{j,m} |S^A_{i_m}S^R_{ji_m}|} ,} 
		\label{momeqn1CR2Atemp} %\\
	\end{eqnarray}

and to the constant coefficient \(C_{R2R}\) as
	\begin{eqnarray}
		\displaystyle {C_{R2A}} &=&
		\displaystyle {\frac{1}{\Sigma_{j,k,l} |S^R_{ik_l}S^R_{jk_l}|}.} 
		\label{momeqn1CR2A} %\\
	\end{eqnarray}

Thus, the total gains agent \(i\) accrues from its participation in the reputation system are:
	\begin{eqnarray}
		\displaystyle {\Delta CP_i} &=&
		\displaystyle {\frac{\Sigma_{j,m} S^A_{i_m}S^R_{ji_m}}{\Sigma_{j,m} |S^A_{i_m}S^R_{ji_m}|} + \frac{\Sigma_{j,k,l} S^R_{ik_l}S^R_{jk_l}}{\Sigma_{j,k,l}|S^R_{ik_l}S^R_{jk_l}|}.} 
		\label{TotalGainsAi} %\\
	\end{eqnarray}

Given that \(\Delta CP_i\) can become negative due to losses, it is imperative to ensure that the cumulative result of the formulas mentioned earlier always remains less than the total assets possessed by agent \(i\). To achieve this, we introduce an added assumption: within the same category, whether actions or ratings, credit points can be internally redistributed or restaked. However, credit points designated for actions cannot be reallocated for ratings, and vice versa. This means that credit points staked for one category can be used within that category's respective behaviours\footnote{Here, the term "behaviour" is used to encompass all three types: action, rating, and trading}, but not beyond. Furthermore, for each agent, the maximum loss within a category cannot exceed the credit points allocated for that specific category. This stipulation implies that every agent must earmark a specific portion of their credit for each staking category from the outset, and these cannot be liquidated until they are unstaked. In more advanced models, this foundational assumption can be replaced with intricate mechanisms that more closely mirror real-world scenarios, ensuring that total losses never surpass assets, even though total gains might exceed them.

\subsection{Distribution of Agent Actions} \label{Distribution of Agent Actions}

For the reputation system, we assume that the action of various agents conforms to a normal distribution \(N(\mu, \sigma^2)\). This is characterized by its average action level, represented by \(\mu\), and its variance \(\sigma^2\). The probability density function of this distribution is:

\begin{equation}
f(x) = \frac{1}{\sigma\sqrt{2\pi}} e^{-\frac{1}{2}\left(\frac{x-\mu}{\sigma}\right)^2},
\end{equation}
where \(x \in (-\infty, \infty)\).

Now, for a specific agent \(i\), we assume that its actions across various activities follow an individual normal distribution. This distribution captures the variability of \(i\)'s actions over time or in different contexts, as opposed to the variability observed among different agents. We represent this normal distribution as \(N(\mu_i, {\sigma_i}^2)\), where it is characterized by its mean action level \(\mu _i\) and its variance \({\sigma_i}^2\). The variance can be assigned a value smaller than that observed across different agents if it more accurately reflects the real-world scenario. The corresponding probability density function for this agent is:

\begin{equation}\label{muSDi}
f_i(x_i) = \frac{1}{\sigma _i\sqrt{2\pi}} e^{-\frac{1}{2}\left(\frac{x_i-\mu _i}{\sigma _i}\right)^2},
\end{equation}
where, once again, \(x_i \in (-\infty, \infty)\).

Although this is an infinite interval, its cumulative distribution function (CDF) is:

\begin{equation}\label{CDF}
F_i(x_i) = \frac{1}{2} \left[ 1 + \text{erf} \left( \frac{x_i - \mu_i}{\sigma_i\sqrt{2}} \right) \right],
\end{equation}
where the error function is defined as:

\begin{equation}
\text{erf}(x_i) = \frac{2}{\sqrt{\pi}} \int_0^{x_i} e^{-t^2} \, dt.
\end{equation}

The CDF lies within the finite interval [0,1]. For \(F(x_i)_i(x_i=0)\), representing the cumulative distribution from \((- \infty, 0)\), it indicates the likelihood of the agent showing negative action on average. Hence, CDF can be used to simulate as the cumulative probability of negative ratings from others in general.\footnote{Here, the term "action level" is used as a relative measure to describe the positive or negative inclination of an agent's actions. A higher action level implies a positive action, while a lower one suggests the opposite. In addition, we do not distinguish the intrinsic or intentional level of action of an agent and the cumulative probability of negative ratings from others since the rating is observable but "intrinsic or intentional" level of action is not.}

\subsection{Action Analysis in Reputation Systems} \label{Action Analysis in Reputation Systems}

Considering a reputation system where agents rate each other based on action, we posit that an agent's rating can either be positive or negative. Each agent's action and rating can influence the gains or losses for others. Thus, the total gains agent \(i\) accrues from its participation in the reputation system are given by formula (\ref{TotalGainsAi}).

Given that the inherent nature of an individual's action is unobservable, we do not differentiate between others' ratings and the actual nature of one's action. Instead, we interpret \( F(x)_i^A(x=0) \) as the likelihood of receiving negative ratings on its action on average.

Let us denote the probability of agent \(i\) providing a positive rating as \(P_+\) and that of negative rating as \(P_- \). The joint probabilities of rating interactions between \(i\) and \(j\) under different scenarios are:\\
1. \(i\) and \(j\) both provide positive rating: \( P_{++} = P_+ \times P_+ \).\\
2. \(i\) and \(j\) both provide negative rating: \( P_{--} = P_- \times P_- \).\\
3. \(i\) provides positive rating, and \(j\) provides negative rating: \( P_{+-} = P_+ \times P_- \).\\
4. \(i\) provides negative rating, and \(j\) provides positive rating: \( P_{-+} = P_- \times P_+ \).

The chances that rating from two agents align in sentiment (either both positive or both negative) is:

\begin{eqnarray}
P_{\text{align}} = P_{++} + P_{--} = P_+^2 + P_-^2.
\end{eqnarray}

To ascertain whether the majority of agents will benefit from such a system without specially designed favourable mechanism, we examine the relative magnitude of probabilities for congruent rating (both positive or both negative) against incongruent rating. Specifically, we need to verify:

\begin{equation}
P_+^2 + P_-^2 \geq 2 \times P_+ \times P_-.
\label{++--+--+}
\end{equation}

Breaking it down and reordering:

\begin{equation}
(P_+ - P_-)^2 \geq 0.
\end{equation}

By the fundamental properties of real numbers, the square of any real number is always non-negative. This implies that the combined probability of agents giving congruent rating (either both positive or both negative) will always be at least as high as that of giving incongruent rating. Thus, when agents' rating aligns, the majority in the system stands to benefit. This dynamic promotes a system where congruent ratings are more prevalent than incongruent ones, underscoring the collaborative nature of such reputation systems. In theory, this mechanism should encourage agents to participate actively in ratings. We can use Monte Carlo simulations to determine whether this is noticeable in practice.

\section{Advanced Models for Real-World Scenarios} \label{Advanced Model}

As we venture deeper into the intricacies of adapting blockchain models for practical, real-world applications, this section elucidates several advanced modeling techniques tailored to address real-life complexities. We commence by understanding the strategy of \textbf{Allowing Agents to Skip Rounds}, which presents flexibility in participation and caters to intermittent engagements. Transitioning to the realm of consumer interactions, we present \textbf{Non-Random Consumer Selection}, detailing methodologies for targeted and non-arbitrary engagement. Recognizing the dynamic nature of real-world systems, we then explore an \textbf{Adaptive Model with Learning and Adjustment}, ensuring our model remains agile, responsive, and self-evolving. Diving into motivational structures, the \textbf{Contribution Incentive Mechanisms} subsection delineates strategic rewards to galvanize proactive participation. Lastly, foreseeing the future of decentralized systems, we examine the \textbf{Integration with Decentralized Autonomous Organizations}, highlighting the fusion of our advanced models with self-governing, decentralized entities. 

\subsection{Allowing Agents to Skip Rounds} \label{sec:skip_rounds}

In real-life scenarios, it is not a given that every individual will interact with or evaluate every other individual at all times. A parallel situation can naturally be anticipated once the reputation system is operational. To enhance the model's reflection of reality, we will design it to allow agents the flexibility to skip rounds. This implies that in each round, a certain proportion of agents may not undertake any actions, and a proportion may refrain from providing any ratings. In all the aforementioned formulae, assigning a value of zero to the staking amount of credit points signifies that the agent is abstaining from participating in the current round. To state explicitly, \(S^A_{i,t}=0\) denotes that agent \(i\) refrains from actions in round \(t,\) and \(S^R_{j,t'}=0\) signifies that agent \(j\) abstains from providing ratings in round \(t'\).

\subsection{Non-Random Consumer Selection} \label{sec:consumer_selection}

As described in Section \ref{Connecting the real world}, when agents (C1) stake more credit points, they are more likely to respond actively to ratings due to the potential gains or losses brought about by the ratings from others. Logically speaking, agents who are willing to stake more credit points are more likely to meet the needs and expectations of consumers. Owing to the transparency of staking information, consumers tend to prefer producers who have staked more credit points due to the nature of risk aversion \cite{Menezes1970, Friedman1974, Pratt1978, Morin1983, Guiso2004}. To initially implement this model, which integrates non-random consumer selection, we can adopt a method wherein, in each round, agents who stake fewer credit points not only have a higher probability of being bypassed for their actions, but also increase the likelihood of being bypassed by other agents for evaluations.

\subsection{Adaptive Model with Learning and Adjustment} \label{Adaptive Model with Learning and Adjustment}

%\subsection{Agent Action and Staking Dynamics in a Reputation System}\label{Agent Action and Staking Dynamics in a Reputation System}

In the intricate realm of reputation systems, the actions of agents are influenced by both their past experiences and the ratings received from their peers. This section delves into the complex interplay of these factors, with a focus on the dynamics of staking in response to observed outcomes.

\subsubsection{Self-Staking and Adjustment Based on Ratings}\label{Self-Staking and Adjustment Based on Ratings}

In the development of our reputation system, agents are characterized by their actions and ratings across different rounds of interaction, each denoting a time step in the system. Agents engage in actions, staking a certain amount of credit points to signal their confidence or commitment. However, not every agent is obligated to perform an action or provide rating in every round, allowing for a more general and flexible model representation. The rating provided by other participants in the system can be either positive or negative and subsequently influences the credit point balance of the agent in focus. This change in credit points is modulated by the initial stake of the agent and the stake accompanying the rating.

To facilitate computation and Monte Carlo simulation, the model introduces time indices representing different rounds of interactions. For an agent \(i\) in round \(t\), we redefine the notation as follows: Let \(S^A_{i,t}\) represent the credit points staked on action \(m\) in round \(t\) when acting as a producer or consumer, which is always a positive value. Meanwhile, \(S^R_{ik,t}\) represents the credit points staked on rating the action \(l\) of another agent \(k\) in round \(t\). This value can be positive if the ratings align in direction, and negative if they are opposite. Here, \(m\) and \(l\) denote different actions within the round \(t\), and agents may not necessarily undertake an action or provide a rating in every round.

\begin{equation}
\Delta CP^A_{i,t} = f(S^A_{i,t}, S^R_{ji,t})
\end{equation}

Where:
\begin{itemize}
    \item \(\Delta CP^A_{i,t}\) represents the change in credit points for agent \(i\) at time \(t\).
    \item \(S^A_{i,t}\) is the staking amount by agent \(i\) at time \(t\).
    \item \(S^R_{ji,t}\) denotes the rating (along with its stake) from another agent \(j\) at time \(t\).
    \item \(f\) is a function that determines the change in credit points based on the stake and rating.
\end{itemize}

In the subsequent round, the agent adjusts their action based staking on the outcomes they observed:

\begin{equation}
S^A_{i,t+1} = S^A_{i,t} + \alpha_L \times \text{sign}(\Delta CP(A)_{i,t}) \times \min(|\Delta CP^A_{i,t}|, \beta \times \left|S^A_{i,t}\right|, \beta \times \left| 1-S^A_{i,t}\right|)~,
\end{equation}

where \(\alpha_L<1\) is the learning rate parameter, which ranges in [0,1] (with 0\% representing the non-learning case), and the minimum function, along with the constant parameter \(\beta\) (where \(\beta < 1\)), ensures that the staking rate remains non-negative and meaningful in the range of [0,1].

Following the equation, the positive or negative sign of \(\alpha_L \times \text{sign}(\Delta CP(A)_{i,t}) \times \min(|\Delta CP(A)_{i,t}|, \beta \times \left|S^A_{i,t}\right|, \beta \times \left| 1-S^A_{i,t}\right|)\) provides insights into the adaptive nature of agents’ staking behavior. Positive outcomes lead agents to gradually increase their staking on actions, amplifying their gains, while agents experiencing negative outcomes, indicative of a higher likelihood of receiving negative ratings, decrease their action staking to limit losses. Over multiple rounds, the number of agents with positive returns increases in high-staking areas, and the number with negative returns decreases, illustrating a natural evolution in the free market where the most proficient individuals specialize in tasks they excel at, and others withdraw from the domain. The exact number of rounds required for significant differences to emerge will be examined through subsequent Monte Carlo simulations.

\subsubsection{Staking on Others and Adjustment Based on Outcomes}\label{Staking on Others and Adjustment Based on Outcomes}

Agents also evaluate the actions of their peers, staking credit points on their evaluations. The magnitude of their stake amplifies the gains or losses they incur, contingent on the congruence of their evaluations with the consensus.

For an evaluation by agent \(i\) on agent \(k\)'s action at time \(t\), the change in credit points is represented as:

\begin{equation}
\Delta CP(R)_{i,t} = g(S^R_{ik,t}, S^R_{jk,t})
\end{equation}

Where:
\begin{itemize}
    \item \(\Delta CP(R)_{i,t}\) denotes the change in credit points for agent \(i\) when evaluating other agents at time \(t\).
    \item \(S^R_{ik,t}\) is the staking rate of agent \(i\) on the action of agent \(k\) at time \(t\).
    \item \(S^R_{jk,t}\) represents the staking rate of a different agent \(j\) on the action of agent \(k\) at time \(t\).
    \item \(g\) is a function determining the change in credit points based on the stake and the evaluated action.
\end{itemize}

To clarify, in the above formula, \(j\) and \(k\) represent indices running through all the agents in the system.

In subsequent rounds, agents adjust their staking rates for ratings based on the outcomes observed. Here, we use \(S^RT_{i,t+1}\) to represent a temporary result before considering the total staking rate, which also includes that for actions:

\begin{equation}
S^RT_{i,t+1} = S^R_{i,t} + \alpha_L \times \text{sign}(\Delta CP(R)_{i,t}) \times \min \left( \left| \Delta CP(R)_{i,t} \right|, \beta \times  \left| S^R_{i,t}\right|, \beta \times  \left| 1 - S^R_{i,t}\right| \right)
\end{equation}

Subsequently, to ensure that the total staking does not exceed 100\%, additional constraints are applied:

\begin{equation}
S^R_{i,t+1} = \min \left( 1 - S^A_{i,t+1}, S^RT_{i,t+1} \right)
\end{equation}

where \(\alpha_L < 1\) is the learning rate parameter, which ranges in [0,1] (with 0\% representing the non-learning case), and \(\beta < 1\) ensures that the staking rate remains non-negative and within the [0,1] range.

This dual-layered rating and adjustment mechanism induces a dynamic evolution of staking actions in the system, thereby ensuring adaptability and responsiveness to the continuously shifting landscape of agent actions and evaluations.

In summary, understanding and modeling the dynamics of actions and ratings is crucial in predicting the evolution of staking systems, and our approach, combining psychological insights, economic reasoning, and rigorous mathematical modeling, offers a comprehensive view of the potential outcomes in such systems.

\subsection{Contribution Incentive Mechanisms} \label{sec:Contribution Incentive Mechanisms}

Enhancing the robustness and sustainability of the reputation system requires more than just leveraging its intrinsic advantages, such as more authentic ratings and demonstrated integrity. It is equally crucial to incentivize those participants who bring significant positive externalities to the platform.

A positive externality \cite{Dahlman1979, Bergh2010} arises when the actions of an individual or firm confer benefits on others without being compensated for them. A standard illustration from economics is the realm of education: individuals who invest in their education not only gain personal advantages but also indirectly bestow broader societal benefits, like spurring productivity and innovation.

Within our reputation system's framework, such positive externalities could emanate from:

\begin{itemize}
    \item \textbf{Direct Investment:} Participants contributing directly to the platform aid its development and expansion. Such contributions can manifest in diverse forms, whether financial resources, technical know-how, or other crucial assets. For instance, backers financing the inception of novel features can amplify the platform's capabilities, thereby magnetizing a larger user base. This prospect aligns with the evident interest of traditional financial entities that are keen on supporting innovative blockchain ventures or other forms of alternative finance \cite{Andrianto2017, Singh2022}.
    \item \textbf{Liquidity Provision:} By ensuring seamless transactions and bolstering market efficiency, liquidity providers hold paramount importance. In the sphere of decentralized finance (DeFi), they are indispensable, pouring assets into liquidity pools to facilitate smoother trade operations while reaping returns for their contributions. Furthermore, it's worth considering tailored incentive schemes for liquidity providers within centralized cryptocurrency exchanges, which often resonate more with users less versed in blockchain intricacies. Established financial institutions, as well as individual economists, advisors, and analysts, are progressively showing interest in exploring or trading in cryptocurrencies \cite{KlagesMundt2022, Tang2023, Alexander2023}.
    \item \textbf{Community Engagement:} An involved community is the backbone of a platform's success, as they amplify its virtues, actively partake in governance, and regularly offer feedback. A fervent user community can catalyze widespread adoption, inspire collaboration, and engender a shared sense of purpose among its members.
\end{itemize}

\subsubsection{Modeling Externality Incentives}

To incorporate externality incentives into the model, we introduce an additional parameter, \(\gamma\), representing the impact of positive externalities on the reputation of agents. The total reputation of an agent \(i\) in round \(t\) can be expressed as:

\begin{equation} \label{Externality}
CP_{i,t} = CP_{i,t-1} + \Delta CP^A_{i,t} + \Delta CP^R_{i,t} + \gamma CP^E_{i,t}
\end{equation} 

where \( CP_{i,t} \) is the credit points of agent \(i\) at time \(t\), \( \Delta CP^A_{i,t} \) is the credit point change w.r.t. actions of the agent, \( \Delta CP^R_{i,t} \) is the  credit point change w.r.t. ratings of the agent, and \( CP^E_{i,t} \) represents the positive externality contributed by the agent.

\subsection{Integration with Decentralized Autonomous Organizations} \label{Integration with Decentralized Autonomous Organizations}

Once both the reputation system and the contribution incentive mechanisms have been simulated and rigorously tested without any issues, further development of the blockchain can be driven by the route of a DAO (Decentralized Autonomous Organizations) \cite{Wang2019, Diallo2018, ElFaqir2020, DuPont2017}. In a Decentralized Autonomous Organization (DAO) model, user engagement is enhanced in two significant capacities. Firstly, as users, they execute actions, conduct trades, and cast ratings. Secondly, they serve as governors, casting votes on rule changes and deciding the overall direction of the blockchain. Their dual role ensures alignment with the reputation system's success, as the platform's growth and sustainability directly impact their benefits.

Drawing inspiration from existing blockchain projects can expedite the integration of DAO principles into our reputation system. \emph{Aragon}, for instance, specializes in providing tools and a platform for decentralized organizations to operate \cite{aragonwhitepaper2019,aragonwhitepaper2017}. It offers a suite of applications and services that streamline the creation and management of decentralized autonomous organizations, thereby removing many of the traditional barriers to entry. On the other hand, \emph{Tezos} represents a self-amending blockchain where decisions about protocol upgrades are made through on-chain governance, relying on stakeholders' voting \cite{allombert2019, perez2020}. Stakeholders can propose, select, or vote on amendments, allowing the blockchain to evolve and adapt without the need for a hard fork. Adapting elements from Aragon's specialized DAO solutions and Tezos' self-evolving structure can help our reputation blockchain achieve the desired autonomy and adaptability, ensuring long-term relevance and resilience.

\section{Monte Carlo Simulations} \label{sec:Monte_Carlo}

Having meticulously constructed our models in preceding sections, it's now imperative to subject them to rigorous simulations and discern whether their performance aligns with our expectations. Central to our simulation methodology is the renowned Monte Carlo technique. We begin our journey with a \textbf{Brief Introduction to Monte Carlo Simulation}, setting the stage for newcomers and grounding our approaches in a theoretical foundation. We then delve into the \textbf{General Settings for Monte Carlo Simulation of the Reputation System}, laying out the parameters and configurations vital for our experiments. Two contrasting initial credit distribution scenarios are subsequently examined: one where \textbf{Scenario of Credit Points Uniformly-Distributed among Agents Initially}, and another highlighting a \textbf{Scenario of Credit Points Initially Distributed in Power-Law among Agents}. As we venture deeper into the simulation results, our focus bifurcates into two segments: the \textbf{Results and Interpretations of Monte Carlo Simulations for the Basic Model} and the \textbf{Results and Interpretations of Monte Carlo Simulations for Advanced Models}, each shedding light on their respective model outcomes and drawing comparisons where necessary.

\subsection{Brief Introduction to Monte Carlo Simulation} \label{sec:Brief_Introduction_to_Monte_Carlo_Simulation}

Monte Carlo Simulation (MCS) \cite{MetropolisUlam1949} is a powerful statistical technique that enables a probabilistic approach to comprehending complex systems. Essentially, MCS employs random sampling and statistical analysis to approximate solutions to mathematical challenges.

\subsubsection{Basic Principles}

The central concept of MCS is simple: by executing a large enough number of random simulations or "trials", one can derive an approximation of a desired quantity. Mathematically, let's take a random variable \(X\) with an expected value \(E[X]\). The Monte Carlo estimate of \(E[X]\) is:

\begin{equation} \label{MonteCarloEx}
E[X] \approx \frac{1}{N_{sim}} \sum_{i=1}^{N_{sim}} X_i
\end{equation}

where \(X_i\) are independent samples drawn from the distribution of \(X\), and \(N_{sim}\) represents the number of simulations.

\subsubsection{Application to Structural Reliability}

In structural engineering, from which MCS has its origins \cite{Luo2022}, MCS has become an essential tool for gauging the reliability of structures. Let's consider a limit state function (LSF) \(G(x)\) that delineates the boundary between failure and successful states of a structure. A typical representation is:

\begin{equation} \label{MonteCarloGx}
G(x) = Rand_R - Rand_S
\end{equation}  
Here, \(Rand_R\) signifies the resistance of the structure (a random variable influenced by factors like material properties and geometry), while \(Rand_S\) corresponds to the applied load or stress (another random variable determined by external influences).

The structure's failure probability \(P_f\) can be defined as the probability where \(G(x)\) is less than or equal to zero:
\begin{equation} \label{MonteCarloPf}
P_f = P(G(x) \leq 0)
\end{equation} 

To estimate this failure probability through MCS, one can:

\begin{enumerate}
    \item Randomly sample values for \(Rand_R\) and \(Rand_S\) from their respective probability distributions.
    \item Calculate the value of \(G(x)\).
    \item Check if \(G(x) \leq 0\) (indicating a failure).
    \item Repeat the above for a large number \(N_{sim}\) of simulations.
    \item Estimate the failure probability as:
\end{enumerate}

\begin{equation} \label{MonteCarloFailure}
P_f \approx \frac{\text{Number of Failures}}{N_{sim}}
\end{equation}

Upon determining \(P_f\), a reliability index \( \beta \) can be formulated using:
\begin{equation} \label{MonteCarlobeta}
\beta = -\Phi^{-1}(P_f)
\end{equation}

where \( \Phi^{-1}(.) \) represents the inverse of the standard normal cumulative distribution function. A higher \( \beta \) value signifies increased reliability.

\subsubsection{Limitations and Considerations}

Despite its versatility, MCS has inherent challenges. For intricate problems or situations with extremely low failure probabilities, an enormous amount of simulations might be essential for achieving adequate accuracy. Moreover, selecting appropriate probability distributions for input random variables (like \(Rand_R\) and \(Rand_S\)) is vital, as they directly impact the outcomes.

\subsection{General Settings for Monte Carlo Simulation of the Reputation System} \label{sec:Preparations_Monte_Carlo}

Within complex reputation systems, particularly those governed by mechanisms like 'staked credit points', the challenge lies not just in constructing a robust theoretical framework, but also in validating its viability in diverse real-world scenarios. Here, the Monte Carlo simulation stands as a quintessential tool, bridging the often-gaping chasm between theoretical robustness and empirical efficacy. By simulating numerous rounds of agent actions, rating, and adjustments, we can gain insights into the system's stability, the distribution of credit points, and the effectiveness of the staking mechanism in promoting desired actions.

The nature of staking, especially with non-traditional metrics like 'credit points', introduces layers of variability and potential strategic actions by participants. The Monte Carlo simulation, with its inherent stochastic modeling, enables a comprehensive exploration of these actions under myriad conditions. By running numerous simulations, each representing a possible state of the world, it unveils vulnerabilities, strengths, and unforeseen outcomes tied to the system.

This systematic exploration also establishes a rating loop. Findings from the simulation can guide refinements in the underlying model, ensuring that the reputation system remains adaptive and resilient. For instance, should simulation results indicate certain strategies where producers might overly centralize staking, the theoretical constructs can be adjusted to deter such actions, thereby ensuring a more equitable and effective system, if it is what the model aims for.

To underscore its significance: the Monte Carlo method isn't merely a testing instrument. It's an iterative dialogue between the abstracted world of theory and the unpredictable terrains of practical application. In the context of a reputation system driven by 'staked credit points', this dialogue is crucial, fortifying the model's architecture against real-world exigencies and guaranteeing its longevity and relevance.

The results of the MCSs will be illustrated and discussed in detail in Section \ref{sec:Results_Interpretations_Monte_Carlo} and Section \ref{Results and Interpretations of MCSs}. It's worth noting that some anomalies were observed within the results of our MCSs. These outlier data points can be attributed to the inherent randomness and stochastic nature of the simulation method. Monte Carlo, by its design, employs random sampling to obtain numerical results for problems that might be deterministic in principle. While the law of large numbers ensures that these simulations converge to the expected value over a vast number of runs, occasional deviations from the expected outcome are not only possible but expected due to this randomness. Detailed examination of these anomalies provides insights into the range of possible outcomes, even if they occur with low probability.

\subsubsection{Division of Staking Rates Between Actions and Ratings}\label{Division of Staking Rates Between Actions and Ratings}

In order to ensure that each agent can independently set their staking ratios for actions and ratings, while maintaining an unbiased overall probability, the initial staking ratios are set as follows. Each agent is first assigned a random value \(SR^A\) which acts as the upper limit for their action staking ratio. This is calculated as:

\begin{equation}\label{StakingRateDivisions}
SR^A = rand(n,1)
\end{equation}

Subsequently, the actual action staking ratio \(S^A\) for each agent is determined by multiplying \(SR^A\) by another random value, ensuring that it is within the limit:

\begin{equation}\label{StakingRateActions}
S^A = SR^A \cdot rand(n,1)
\end{equation}

Similarly, the staking ratio for ratings \(S^R\) is determined by considering the remaining portion from the action staking ratio, thus ensuring the total staking ratio sums up to 1:

\begin{equation} \label{StakingRateRatings}
S^R = (1 - SR^A) \cdot rand(n,1)
\end{equation}

\subsection{Scenario of Credit Points Uniformly-Distributed among Agents Initially}\label{Uniform-Distributed Initial Credit Points}

In the context of our reputation system theoretical model, we employ Monte Carlo simulation to generate synthetic data that adheres to specific statistical properties. For the initial MCS of the reputation system model, we propose an approach where each agent is initially provided with an equal share of credit points. Agents, with their respective credit point allocations, will then randomly determine a staking ratio for various activities within the system. The core intention behind this setup is to examine if such an even distribution combined with a random staking approach can incentivize active participation and promote positive actions among agents.

Every agent \( A_i \) in the system, where \( i \) ranges from 1 to \( n \) (total number of agents), starts with an equal amount of credit points:
\begin{equation} \label{CPinitialUni}
CP_{initial} = \frac{CP_{total}}{n}
\end{equation}

where \( CP_{total} \) is the total amount of credit points in the system.

Once credit points are distributed, each agent determines a staking ratio for their activities. This ratio \( S^A_i \) and \( S^R_i \) are random values between 0 and 1 and signifies the proportion of their credit point allocation they are willing to stake for a particular activity, action and rating, respectively.

\begin{equation} \label{StakingRateActionActual}
S^A_i \sim \text{Uniform}(0,1)
\end{equation}
\begin{equation} \label{StakingRateRatingActual}
S^R_i \sim \text{Uniform}(0,1)
\end{equation}

With uniform initial credit point distribution and random staking ratios, the system's primary goal is to understand if agents are inclined towards positive participation. Key points of examination include:

\begin{itemize}
    \item \textbf{Active Participation}: Whether agents, having an equal footing at the beginning, are more likely to participate actively.
    
    \item \textbf{Risk and Reward Analysis}: How agents balance their staking decisions when the potential rewards and risks are uncertain.
    
    \item \textbf{Action Dynamics}: If and how agents adjust their staking ratios based on observed outcomes, either from their own actions or from the actions of other agents.
\end{itemize}

This model offers a fresh perspective on agent action when introduced to a reputation system that starts with equality and then introduces randomness in staking decisions. By monitoring the agents' actions under these conditions, we can gain insights into the optimal strategies and adjustments required to foster a more cooperative and engaged environment.

\subsection{Scenario of Credit Points Initially Distributed in Power-Law among Agents}\label{Power-Law-Distributed Initial Credit Point}

One of the fundamental challenges in designing any computational model is selecting the right initial conditions and distributions that mirror the complexity and irregularity of the real world. Empirical studies across numerous domains, ranging from economics to natural phenomena, often demonstrate that certain systems do not follow a simple uniform distribution. Instead, they lean towards distributions characterized by a power law.

The Pareto Principle, colloquially known as the 80-20 rule, is a striking manifestation of power law distributions in socio-economic settings. It suggests that roughly 80\% of the effects come from 20\% of the causes, be it wealth distribution, sales figures, or other phenomena. Rooted in both economic reasoning and mathematical underpinnings, this principle highlights the skewed nature of many real-world distributions.

In light of this, while beginning with a uniform distribution offers a clean, symmetric baseline for our model, it's imperative to also examine the action under power law distributions. By doing so, we can ensure our model's robustness and its capability to generalize to real-world scenarios where inequalities and imbalances are often the norm.

In the random sampling of MCS, one of the primary challenges is the generation of random numbers that adhere to a non-uniform distribution, such as the power-law distribution. The inverse transform method offers an elegant solution to this problem \cite{Fishman2013, Raychaudhuri2008, Devroye1986, Law2000, Rubinstein2007}.

\subsubsection{Inverse Transformation Method}
\label{subsec:InverseTransformation}

The inverse transformation method is a fundamental technique for generating random samples from a specified distribution. By utilizing the inverse of the cumulative distribution function (CDF) — which is derived from the probability density function (PDF) for continuous variables and the probability mass function (PMF) for discrete variables — this method can transform uniformly distributed random numbers, typically from the interval (0, 1), into values consistent with the target distribution. Referring to the CDF formula \ref{CDF} in Section \ref{Distribution of Agent Actions}, we categorize the application of this method into two scenarios: continuous and discrete.

\textbf{Continuous Case:}
Let \(X_c\) be a continuous random variable with PDF denoted as \(f_c(x)\). The CDF of \(X_c\) is represented as \(F_c(x)\). Assuming \(F_c(x)\) is continuous and strictly increases, its inverse is denoted by \(F_c^{-1}(u)\) or the inverse CDF. This inverse function relates to the uniformly distributed random variable \(U\).

To generate a random value from the continuous distribution \(f_c(x)\), the process is:
\begin{enumerate}
    \item Generate \(U \sim U(0,1)\).
    \item Compute \(X_c = F_c^{-1}(U)\).
\end{enumerate}
For the continuous case, the relationship between \(F_c(x)\) and \(F_c^{-1}(u)\) can be represented as:
\begin{equation}
\label{eq:continuousFinv}
u = F_c(x) \implies x = F_c^{-1}(u)
\end{equation}

\textbf{Discrete Case:}
For a discrete random variable \(X_d\), where the PMF is \(p_d(x_i)\), the cumulative PMF is:
\begin{equation}
\label{eq:discreteFinv}
F_d(x) = P(X_d \leq x) = \sum_{x_i \leq x} p_d(x_i)
\end{equation}

To generate a random value from this discrete distribution, identify the smallest integer \(I\) such that \(U \leq F_d(x_I)\). Then, set \(X_d = x_I\).

\subsubsection{Application of the Inverse Transformation Method to Power-law Distributions}
Power-law, also known as Pareto distributions, appear frequently in various natural and societal phenomena. They are characterized by "heavy tails", indicating that tail events (those of low probability) are still notably impactful.

For a power-law distribution with a PDF given by \(p(x) \sim x^{-\alpha}\) where \(\alpha > 1\), its CDF is:
\begin{equation}
\label{eq:powerlawCDF}
F(x) = 1 - \left( \frac{x}{x_{min}} \right)^{1-\alpha}
\end{equation}
Here, \(x_{min}\) signifies the smallest conceivable value of \(x\).

The inverse of this CDF, essential for the inverse transformation method, is:
\begin{equation}
\label{eq:inversecdfPowerlaw}
F^{-1}(u) = x_{min} \left(1 - u\right)^{-\frac{1}{\alpha-1}}
\end{equation}

The inverse transformation method is favored when modeling systems with power-law traits due to:
\begin{itemize}
    \item Its straightforward implementation, exemplified by the analytical form of the inverse CDF in equation \eqref{eq:inversecdfPowerlaw}.
    \item Its aptness at mirroring the unique heavy-tailed nature innate to power-law distributions.
\end{itemize}

However, for distributions lacking a defined inverse CDF, the method might be less efficacious. Such scenarios would require iterative numerical procedures, potentially augmenting computational costs and introducing potential inaccuracies.

In summation, the inverse transform method adeptly translates uniformly distributed random numbers to ones adhering to a non-uniform distribution. In our reputation system model, this methodology proved invaluable for generating synthetic data that mirrored a power-law distribution, affirming the model's authenticity and precision.

\subsection{Results and Interpretations of Monte Carlo Simulations for the Basic Model} \label{sec:Results_Interpretations_Monte_Carlo}

Following the preparations outlined in Section \ref{sec:Preparations_Monte_Carlo}, we seamlessly executed MCSs on the basic model. The outcomes aligned perfectly with our expectations generated from Section \ref{Basic Model}.

\subsubsection{Credit Point Dynamics based on Agent Actions} \label{subsec:CreditDynamics}

In our model, the implications of agent actions are distinctly reflected through the benefits or penalties they accrue in the form of credit points. Our MCSs categorically delineate agents into two prominent categories based on the mean values assigned to their actions with a smaller variance than that across agents, as introduced in Formula \ref{muSDi} of Section \ref{Distribution of Agent Actions}. 

\begin{enumerate}
    \item \textbf{Beneficial Actors:} Agents endowed with higher mean action values in the simulations invariably experience positive returns when evaluated by their peers. The magnitude of their credit point gains is directly proportional to the amount they staked during their actions and the stake amount of the evaluating peers.
    
    \item \textbf{Detrimental Actors:} Conversely, agents assigned lower mean action values incur losses in credit points when evaluated. Similar to the beneficial actors, the quantum of these losses is directly influenced by the agent's staked amount during their actions and the stake of the evaluating counterparts.
\end{enumerate}

To analyze the variation in action levels among agents, we use the variable CDF, as defined in Formula \ref{CDF} of Section \ref{Distribution of Agent Actions}. As the simulation progresses, differences in credit point distributions become more pronounced based on agents' action quality. This leads to a clear trend: agents with consistent positive actions experience an increase in benefits, while those with negative actions face mounting penalties.

\subsubsection{Stake-to-Reward Relationship with respect to Ratings}\label{sec:stake-reward-relation}

Our MCSs, initiated with a uniform distribution of credit points, elucidate the intricate dynamics between agent actions and the progressive distribution of credit points. As the simulation progresses, we observe that the cumulative benefit or loss an agent incurs for each review is primarily determined by two pivotal factors: the sign (positive or negative) of their review for a particular agent, and its alignment or disparity with the majority consensus. Furthermore, the magnitude of these benefits or losses is also proportionally affected by the amount an agent pledges during the review, as well as the pledges made by other reviewers. Essentially, the staked amount acts as a multiplier, amplifying the potential gains or detriments. Therefore, agents will experience varying cumulative gains or detriments over time. These cumulative benefits, or the sum of individual rating outcomes, depend on factors like the magnitude of their own and other raters' stakes, and especially, the degree of alignment of their ratings with the prevailing majority opinion.

In the context of a power law distribution, our conclusions remain strikingly similar. Our simulations, initialized with agents adhering to a power law distribution of credit points, underscore the complex relationship between individual agent actions and the ensuing distribution dynamics. It becomes evident that an agent's cumulative gain or loss from each evaluation is fundamentally governed by the polarity (positive or negative) of their feedback for a given agent and its congruence or divergence from the majority consensus. The cumulative benefits, being an aggregate of outcomes from individual ratings, are influenced by various factors: an agent's rating frequency, the quantum of stakes they commit during their evaluations, and predominantly, the extent to which their feedback mirrors the predominant majority viewpoint.

\subsection{Results and Interpretations of Monte Carlo Simulations for Advanced Models} \label{Results and Interpretations of MCSs}

It is crucial to simulate more advanced models that closely resemble real-world scenarios. This step is imperative to understand the dynamics and implications of the proposed system comprehensively. Furthermore, before the formal launch of the reputation system, extensive real-world testing with actual users is of paramount importance to ensure the robustness and reliability of the system. This not only helps in identifying potential challenges but also aids in refining and optimizing the model for practical applications.
\begin{figure}[h!] 
  \centering
  \includegraphics[width=0.6\textwidth]{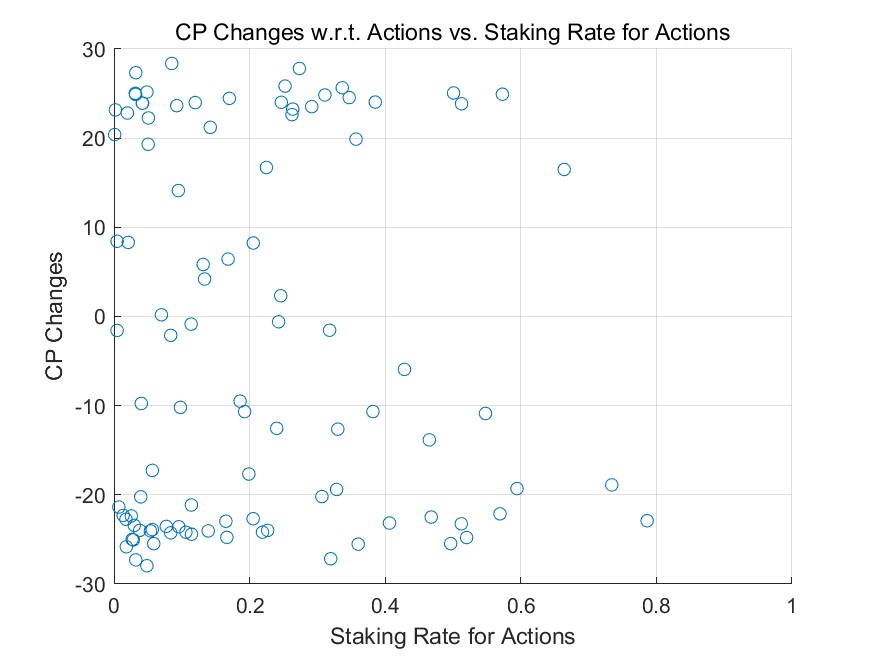} 
  \caption{Non-Learning Model: Changes of Credit Points over Staking Rate for Actions from the uniform initial distribution}
  \label{fig:Non-Learning Actions Uniform}
\end{figure}

\begin{figure}[h!] 
  \centering
  \includegraphics[width=0.6\textwidth]{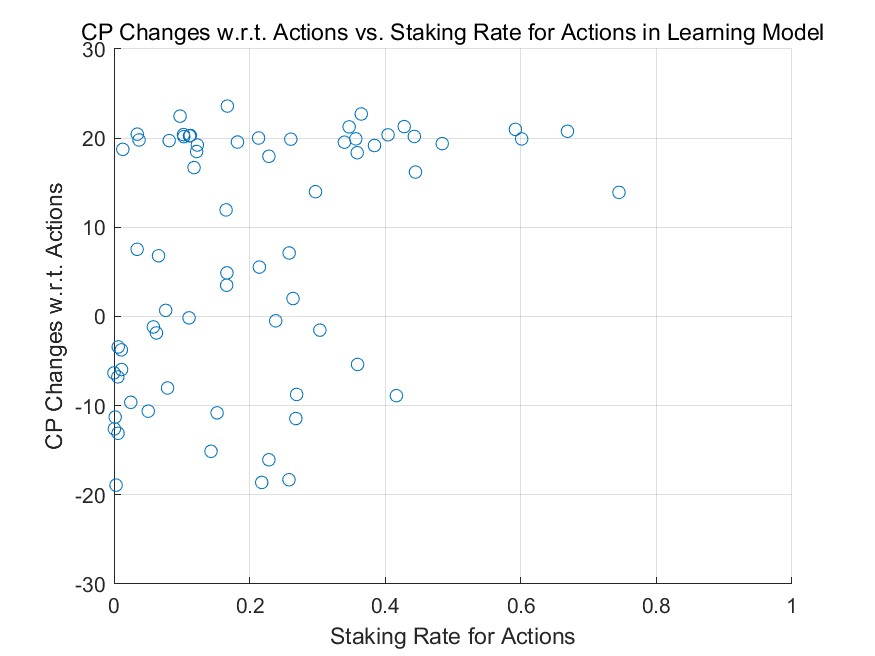} 
  \caption{Learning Model: Changes of Credit Points over Staking Rate for Actions from the uniform initial distribution}
  \label{fig:Learning Actions Uniform}
\end{figure}

\subsubsection{Adaptive Model with Learning and Adjustment} \label{Simulaton Adaptive Model with Learning and Adjustment}

\begin{figure}[h!] 
  \centering
  \includegraphics[width=0.6\textwidth]{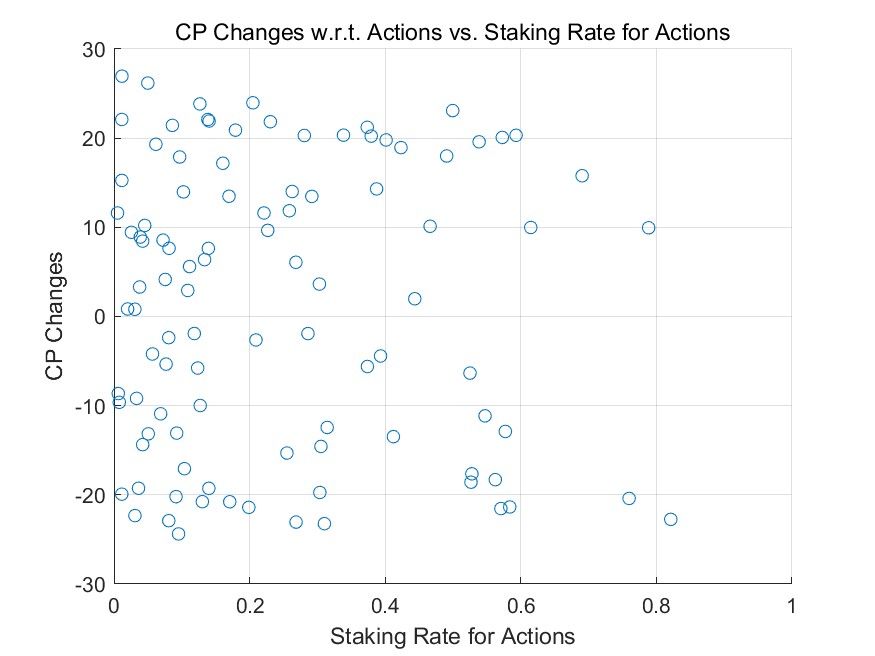}
  \caption{Non-Learning Model: Changes of Credit Points over Staking Rate for Actions from the power-law initial distribution}
  \label{fig:Non-Learning Actions Power}
\end{figure}

\begin{figure}[h!] 
  \centering
  \includegraphics[width=0.6\textwidth]{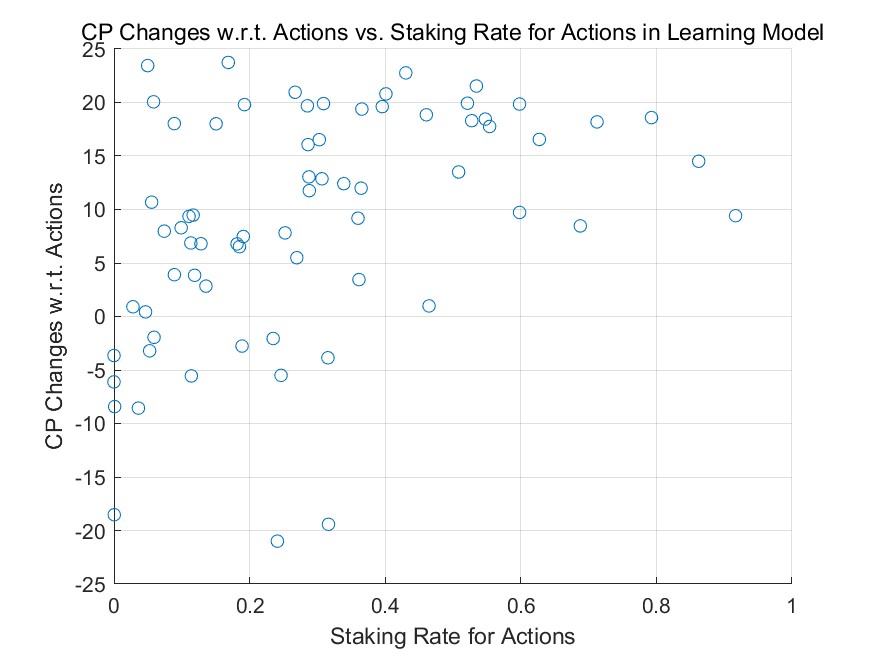} 
  \caption{Learning Model: Changes of Credit Points over Staking Rate for Actions from the power-law initial distribution}
  \label{fig:Learning Actions Power}
\end{figure}

\begin{figure}[h!] 
  \centering
  \includegraphics[width=0.6\textwidth]{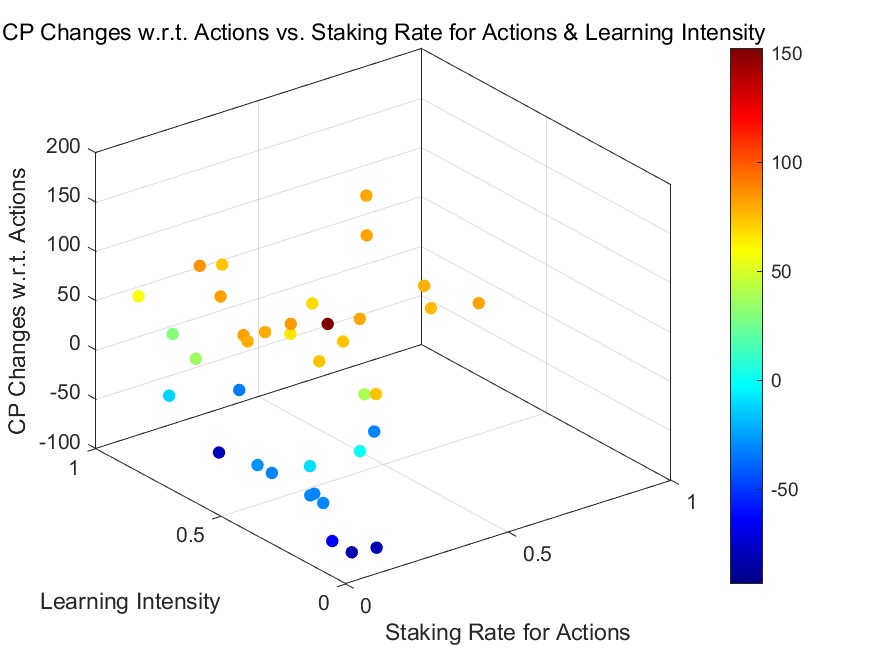} 
  \caption{Changes of Credit Points w.r.t. Actions vs. Staking Rate for Actions \& Random Learning Intensity (Uniform-Initial-Distribution)}
  \label{fig:3D Learning Actions Uniform}
\end{figure}

\begin{figure}[h!] 
  \centering
  \includegraphics[width=0.6\textwidth]{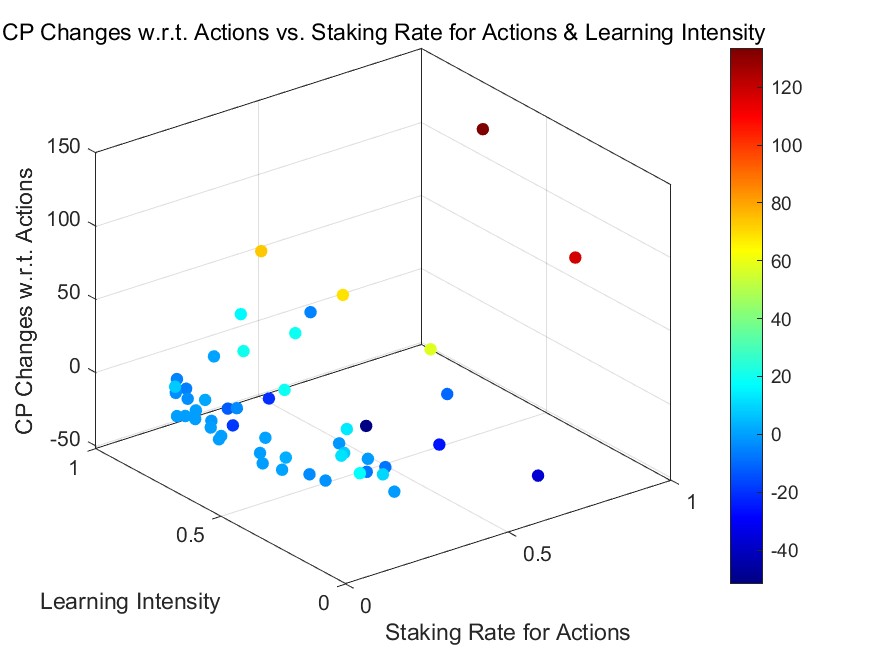} 
  \caption{Changes of Credit Points w.r.t. Actions vs. Staking Rate for Actions \& Random Learning Intensity (Power-Law-Initial-Distribution)}
  \label{fig:3D Learning Actions Power}
\end{figure}

\begin{figure}[h!] 
  \centering
  \includegraphics[width=0.6\textwidth]{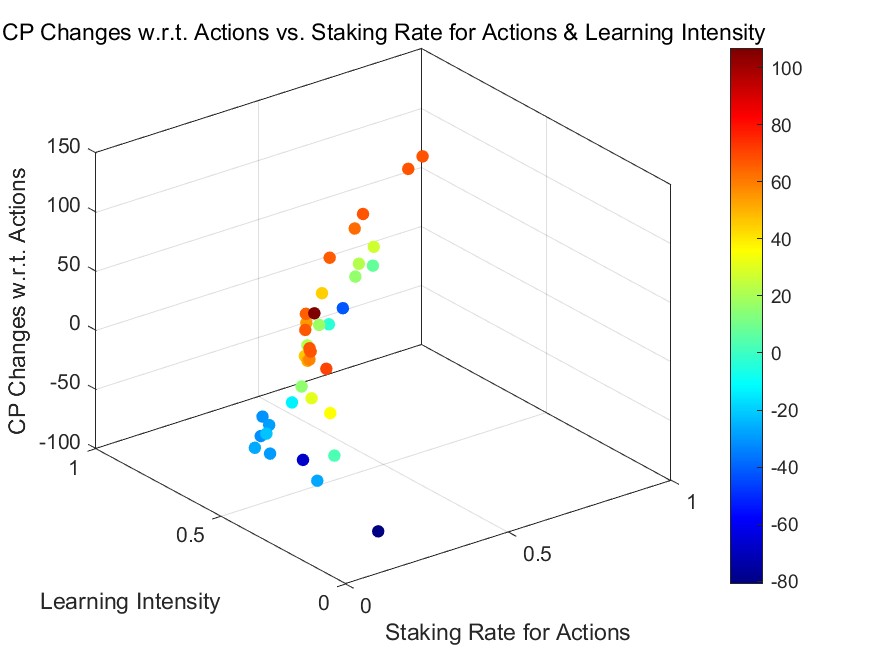} 
  \caption{Changes of Credit Points w.r.t. Actions vs. Staking Rate for Actions \& Positive-Correlated Learning Intensity (Uniform-Initial-Distribution)}
  \label{fig:3D1 Learning Actions Uniform}
\end{figure}

\begin{figure}[h!] 
  \centering
  \includegraphics[width=0.6\textwidth]{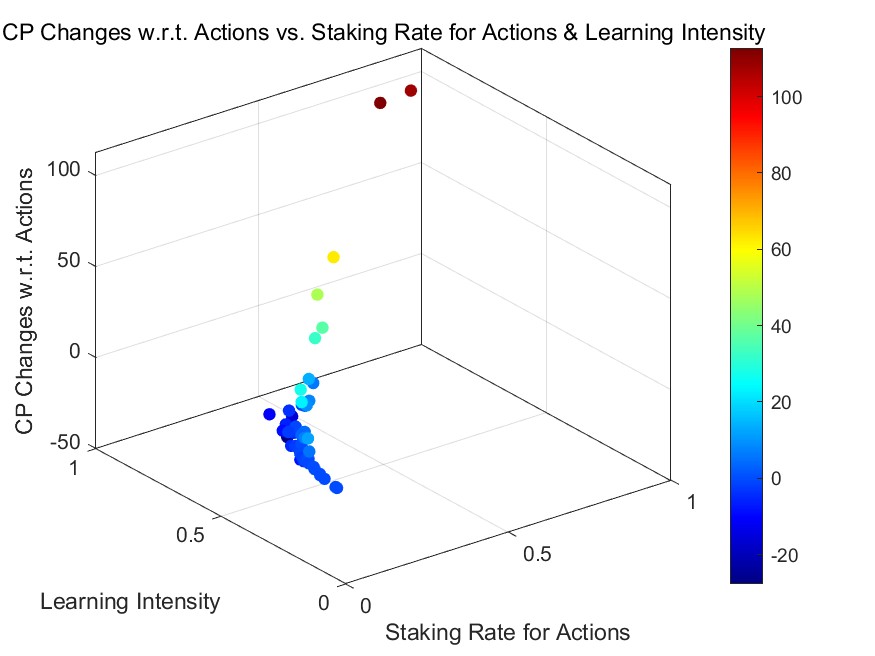} 
  \caption{Changes of Credit Points w.r.t. Actions vs. Staking Rate for Actions \& Positive-Correlated Learning Intensity (Power-Law-Initial-Distribution)}
  \label{fig:3D1 Learning Actions Power}
\end{figure}

Based on the theoretical analysis in Section \ref{sec:consumer_selection}, we discern a marked difference between the adaptive model with learning and adjustment and the non-learning models regarding staking on actions. As rounds progress, agents with positive returns—often acknowledged by their peers—increase their staking, enhancing their rewards. Conversely, agents experiencing negative returns, mainly due to unfavorable ratings, decrease their staking, thereby reducing their losses. The magnitude of these adjustments correlates with learning intensities (denoted by $\alpha_L$, as discussed in Section \ref{Adaptive Model with Learning and Adjustment}) and the return outcomes until the staking rate approaches 0 or 1, at which point intervals become negligible.

This behavior is corroborated by the results of Monte Carlo simulations (MCS), particularly in high-staking regions. A comparative study between the adaptive learning model (see Figure \ref{fig:Learning Actions Uniform} for a uniform initial distribution and Figure \ref{fig:Learning Actions Power} for a power-law initial distribution) and the non-learning model (refer to Figure \ref{fig:Non-Learning Actions Uniform} and Figure \ref{fig:Non-Learning Actions Power}) underscores this trend. Agents with positive outcomes increase their staking in high-staking regions, as evident in the right portions of the figures. In contrast, those with negative results decrease their stakes, gravitating towards the left of the diagrams.

Over multiple rounds, the adaptive model with learning and adjustment demonstrates a rise in agents achieving positive returns in high-staking actions, while the non-learning model shows a decline in such agents. This dynamic resembles the natural evolution of free markets where specialists excel in their domain, echoing the adage, "the most adept individuals handle the most nuanced tasks."

Within the learning model, we utilized Monte Carlo simulations to assess varying learning intensities (denoted by $\alpha_L$) among agents, elaborated further in Section \ref{Adaptive Model with Learning and Adjustment}. We explored two scenarios: uniform learning intensity across agents and individual intensities.

For a uniform $\alpha_L$, the behavior is straightforward. A higher $\alpha_L$ prompts agents with positive outcomes to increase their staking, while those with negative outcomes retract.

We pursued two approaches for individualized learning intensities. Initially, we assigned random values from the range $[0,1]$ to each agent's learning intensity, illustrated in Figure \ref{fig:3D Learning Actions Uniform} for uniform initial credit points and Figure \ref{fig:3D Learning Actions Power} for Power-Law initial credit points. Then, we established a direct correlation between the learning intensity $\alpha_L$ and an agent's staking rate, depicted in Figure \ref{fig:3D1 Learning Actions Uniform} and Figure \ref{fig:3D1 Learning Actions Power}.

For both approaches, the overarching trends are consistent: Agents with higher learning intensities tend to experience amplified credit gains and fewer losses. Influenced by previous outcomes, they adjust their stakes, optimizing their risk-reward balance.

Notably, the patterns are more pronounced in Figure \ref{fig:3D1 Learning Actions Uniform} and Figure \ref{fig:3D1 Learning Actions Power} than in Figure \ref{fig:3D Learning Actions Uniform} and Figure \ref{fig:3D Learning Actions Power}. This distinction arises from the cumulative effect of increasing both staking and learning intensities over rounds, suggesting that credit changes align with these intensities over time.

Moreover, the differences between Uniformly-distributed and Power-Law-distributed initial credit points are consistent with our prior discussions. The role of learning intensity in altering credit points remains uniform across varying initial distributions, with the Power-Law distribution maintaining its long-tail effect.

In contrast to staking on actions, the discrepancy between the adaptive and non-learning models concerning staking on ratings is more subtle. This nuance originates from the complexities of learning from rating consequences. Agents benefit when their ratings align with the majority but face challenges when they differ. This variance somewhat moderates the "Keynesian Beauty Contest" effect \cite{Keynes1936}, where agents anticipate majority ratings rather than providing genuine assessments.

These simulation results resonate with the theoretical insights from Section \ref{Adaptive Model with Learning and Adjustment}. They suggest our blockchain-centric reputation system effectively establishes an on-chain market mechanism, adeptly avoiding the challenges of falsified ratings prevalent in traditional third-party vendor platforms, as highlighted in Section \ref{Introduction}.

\subsubsection{Allowing Agents to Skip Rounds} \label{subsec:simulate_skip_rounds}

We have developed a MCS program that incorporates the mechanism allowing agents to skip rounds. Specifically, in each round, a certain proportion of agents will neither perform any actions nor provide any ratings. Unless otherwise specified, all subsequent simulation results presented will account for this skipping mechanism.

\subsubsection{Simulation of Non-Random Consumer Selection} \label{subsec:simulation_consumer_selection}

\begin{figure}[h!]
  \centering
  \includegraphics[width=0.6\textwidth]{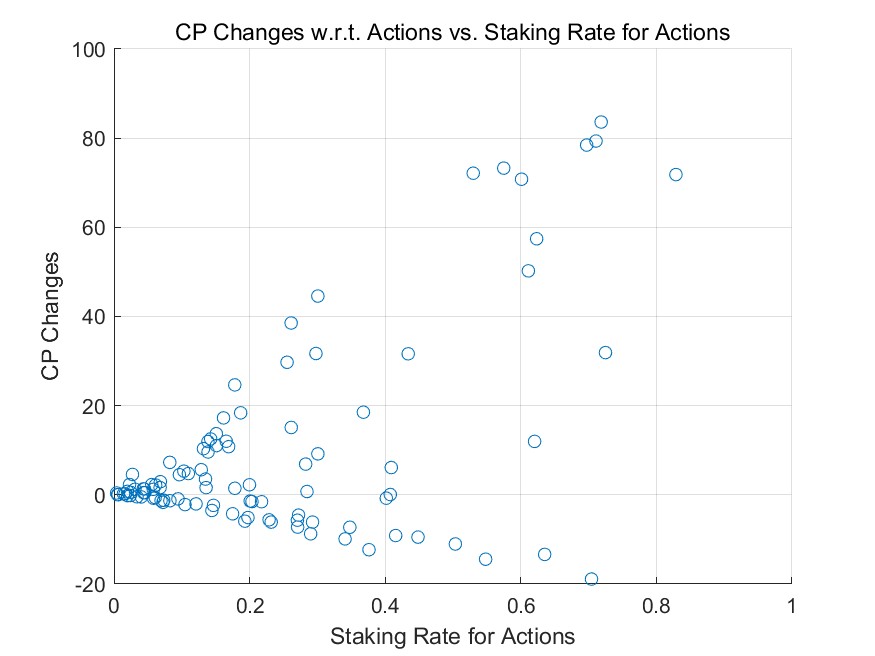} 
  \caption{Non-Learning Model with Consumer Selection: Changes of Credit Points over Staking Rate for Actions from the uniform initial distribution}
  \label{fig:Non-Learning Uniform with Consumer Selection}
\end{figure}

\begin{figure}[h!]
  \centering
  \includegraphics[width=0.6\textwidth]{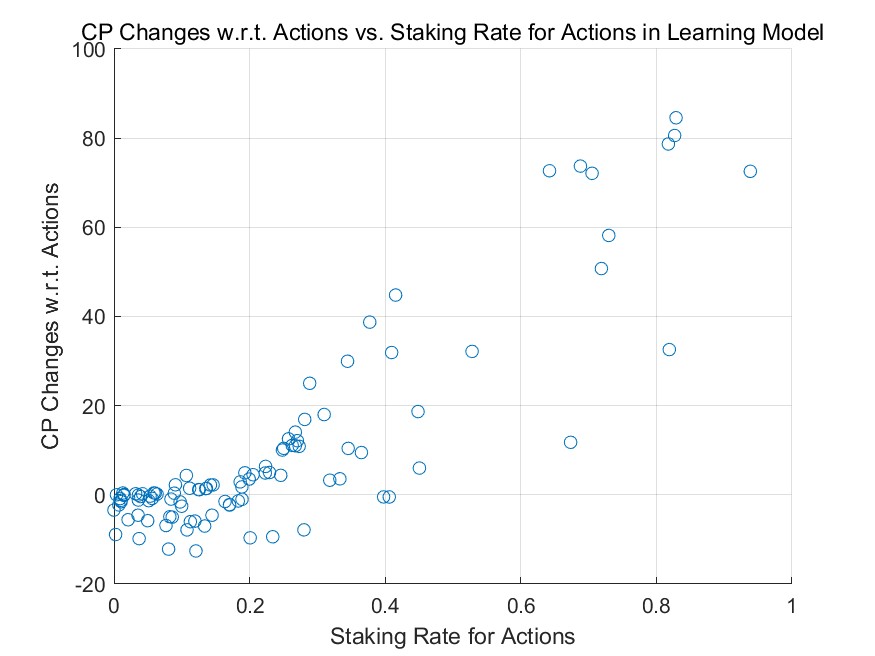} 
  \caption{Learning Model with Consumer Selection: Changes of Credit Points over Staking Rate for Actions from the uniform initial distribution}
  \label{fig:Learning Uniform with Consumer Selection}
\end{figure}

\begin{figure}[h!] 
  \centering
  \includegraphics[width=0.6\textwidth]{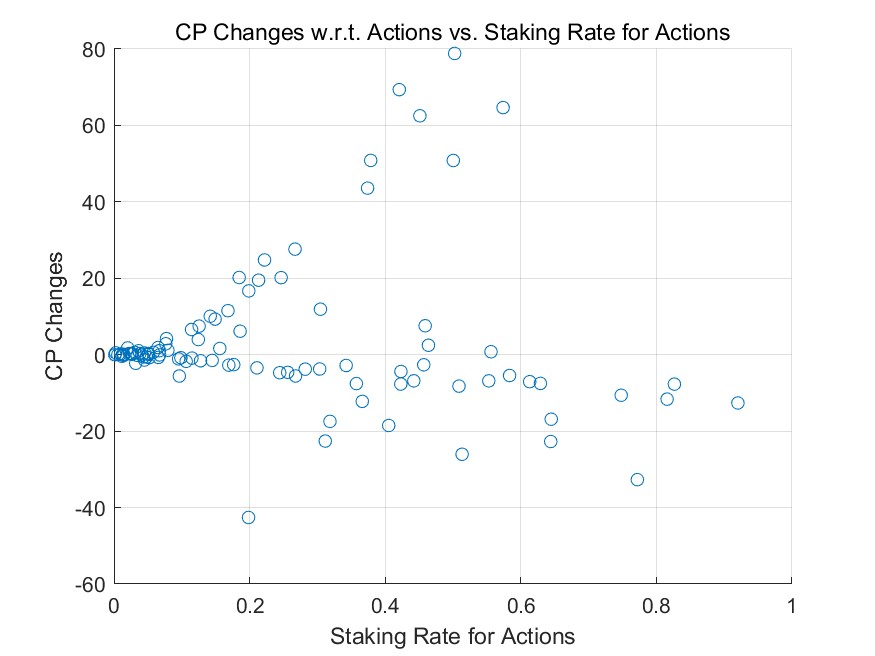} 
  \caption{Non-Learning Model with Consumer Selection: Changes of Credit Points over Staking Rate for Actions from the power-law initial distribution}
  \label{fig:Non-Learning Power with Consumer Selection}
\end{figure}

\begin{figure}[h!] 
  \centering
  \includegraphics[width=0.6\textwidth]{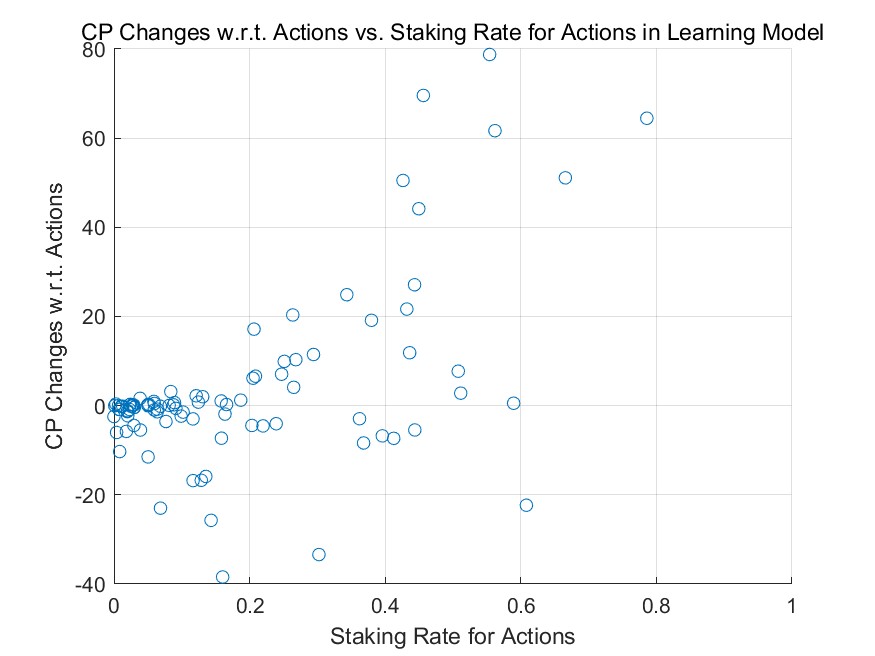} 
  \caption{Learning Model with Consumer Selection: Changes of Credit Points over Staking Rate for Actions from the power-law initial distribution}
  \label{fig:Learning Power with Consumer Selection}
\end{figure}

As elaborated in Section \ref{sec:consumer_selection}, the transparency of staking information naturally inclines consumers to favor producers who have staked a larger amount of credit points. Given that consumers inherently favor provider agents with a higher stake in credit points, it follows that those with minimal credit points staked for actions are more likely to be overlooked and skipped by consumers, in line with the skipping mechanism delineated in Section \ref{sec:skip_rounds} and Section \ref{subsec:simulate_skip_rounds}. We now incorporate this skipping mechanism in simulating non-random consumer selection.

The MCSs elucidate the dynamics of credit points relative to the staking rate for actions, especially when consumer selection mechanisms are incorporated. Starting from a uniform initial distribution, the outcomes for the non-learning model are depicted in Figure \ref{fig:Non-Learning Uniform with Consumer Selection}, while the learning model outcomes are presented in Figure \ref{fig:Learning Uniform with Consumer Selection}. Outcomes derived from the power-law initial distribution are similarly demonstrated in Figure \ref{fig:Non-Learning Power with Consumer Selection} and Figure \ref{fig:Learning Power with Consumer Selection}.

Across these simulations, a clear trend becomes evident: the influence of consumer selection mechanisms results in virtually negligible positive or negative returns for those who stake minimal credit points on actions. The patterns observed in the high staking rate regions, or the right half of the figures, align with previous observations: agents receiving positive returns are inclined to increase their stakes, whereas those incurring negative returns tend to reduce their staking rates, shifting towards the lower staking regions, i.e., the left sections of the figures.

This again tells, tasks tend to be more frequently allocated to individuals who stake a higher number of credit points for their actions. From a national economic standpoint, it is plausible to infer that these agents are more willing to commit to their actions, perhaps due to their specialized skills, or for other reasons. Consequently, they often receive positive ratings. On the flip side, those who are less specialized or consistently receive less favorable reviews find themselves marginalized over time. This evolving landscape resonates with the time-tested belief: "the most specialized individuals handle the most specialized tasks." 

\subsubsection{Contribution Incentive Mechanisms}\label{subsec:contribution}

MCSs with contribution incentive mechanisms indicate that the allocation of credit points for actions and ratings outside the system can dilute the credit points for others. However, the network effect, which brings competitive advantages among peers, can enhance the intrinsic value of each credit point. It is anticipated that if the incentive mechanism is designed appropriately, meaning the increase in credit points outweighs the dilution effect, then overall, everyone in the system stands to benefit.

\section{Conclusions} \label{sec:Conclusions}

%Intrinsic Incentive Mechanisms for Integrity-driven Ratings; Incentive Mechanisms for Actions; Balancing Stakeholders' Shares and Network Effects; The Everlasting Reputation System as a Universal Sustainable Oracle

Building on the insights drawn from our Monte Carlo simulation results, this culminating section distills our findings and teases out the intricate nuances that emerged throughout our study. We commence by probing the bedrock of our model—\textbf{Intrinsic Incentive Mechanisms for Integrity-driven Ratings}—which underscores the spontaneous motivations driving genuine ratings. Progressing further, we then delineate the \textbf{Incentive Mechanisms for Actions}, shedding light on the various levers and pulleys that can be manipulated to promote desired agent behaviors. As a summary to the contribution incentive mechanisms, a crucial balancing act then comes into focus as we navigate the dynamics of \textbf{Balancing Stakeholders' Shares and Network Effects}, emphasizing the delicate equilibrium necessary for a thriving ecosystem. Rounding off our discourse, we elevate our gaze to a holistic vista, capturing the grand vision of the \textbf{The Everlasting Reputation System as a Universal Sustainable Oracle}, which places our system in the broader context of the blockchain oracle landscape.

\subsection{Intrinsic Incentive Mechanisms for Integrity-driven Ratings} \label{subsec:Incentive Mechanisms for Integrity-driven Ratings}

In the realm of online ratings, ensuring integrity-driven feedback is pivotal. Our system's design incorporates a mechanism that penalizes agents whose ratings diverge from the consensus, especially when they have staked a significant amount. This is instrumental in thwarting malicious actors. Vendors offering subpar products or services, contemplating artificially boosting their ratings or undermining their rivals, confront imminent hefty losses. Even if spurious reviews momentarily influence the consensus, integrity-driven reviews, over time, can redress the balance. This may relegate their staked ratings to the minority, inflicting substantial losses.

Concurrently, vendors deliberating on posting negative reviews against competitors with top-notch products or services encounter an analogous predicament. Considering that the majority of consumers are predisposed to furnish positive feedback for high-caliber products or services, a malevolent negative review can swiftly be overshadowed, culminating in substantial losses for the deceitful vendor. This architecture ensures that, viewed through the lens of economic incentives, agents are propelled to rate with an integrity-driven mindset grounded in their firsthand experiences.

Venturing into game theory, while agents may kick off with endeavors to ascertain the predominant sentiment, the game is poised to settle into a state where the lion's share of agents unambiguously rate driven by their integrity-driven experience. This resonates with specific game theoretic models where cyclical guessing games gravitate towards the integrity-driven modus operandi espoused by autonomous participants \cite{Keene2023}.

Such findings not only bolster our theoretical postulations but also accentuate the tangible applicability of our blockchain-anchored reputation system. By synergizing individual motivations with the overarching system mechanics, we unfurl a resilient platform wherein integrity-driven ratings gain momentum, while duplicitous manipulations face retribution.    

\subsection{Incentive Mechanisms for Actions} \label{subsec:Incentive Mechanisms for Actions}

From our Monte Carlo Simulations (MCS) results and interpretations, we've distilled core insights on action-related staking behaviors within a blockchain-based reputation system:

A clear distinction arises when comparing the adaptive model, which includes learning and adjustment mechanisms, to its non-learning counterparts in the context of staking on actions. Agents receiving positive feedback often increase their stakes, aiming to maximize their rewards. Conversely, those encountering negative outcomes tend to reduce their stakes to minimize losses.
    
The trends observed in high-staking scenarios further reinforce these findings. Agents reaping positive benefits amplify their staking commitments, whereas those on the receiving end of negative feedback tend to retract. This pattern is akin to free-market behaviors where tasks naturally shift towards more proficient agents, sidelining the less adept ones.
    
The introduction of features such as the option for agents to skip rounds and the utilization of non-random consumer selection provides deeper insights into staking behaviors. Notably, tasks are predominantly assigned to agents who commit significant credit points, an indicator of their dedication or skill set in specific roles. Agents with higher stakes simplify the selection criteria for consumers, giving them a competitive advantage and attracting newcomers. As a result, our blockchain-based reputation system effectively manages an on-chain marketplace, steering clear of the fake rating problems common in traditional platforms.

In summary, these findings highlight the platform's intrinsic mechanisms for encouraging integrity and cultivating a beneficial relationship between consumers and providers. Importantly, the ability of our platform to counter fake reviews gives it a competitive edge, drawing in users and third-party vendors from competing systems. This positive feedback not only enhances the platform's reputation but also leverages the network effect \cite{Peterson2018, Swann2002}, strengthening its market presence.

\subsection{Balancing Stakeholders' Shares and Network Effects} \label{subsec:balance_incentives_effects}

Through our simulations, which incorporate contribution incentive mechanisms, we elucidate the potential equilibriums in credit point distributions. While external incentives may reduce the shares of existing stakeholders, they can simultaneously enhance the intrinsic value of each credit point via the network effect.

There is a pivotal balance, achievable through judiciously crafted mechanisms and DAO governance, where an optimal outcome that serves all stakeholders' interests can be realized. The aim is to crystallize this equilibrium and possibly enshrine it within a constitutional framework, thus establishing a robust foundation for the reputation system's functionality.

\subsection{The Everlasting Reputation System as a Universal Sustainable Oracle}

The rapid progression of blockchain technology has opened new horizons for decentralized applications. Within this paradigm, the oracle problem---the challenge of securely and reliably transmitting external data to the blockchain---emerges as a significant hurdle. Although Chainlink's decentralized oracle aimed to address this, it faces challenges concerning long-term sustainability and potential dilution of asset value. 

Drawing from the inherent features of blockchain such as transparency and immutability, the everlasting reputation system presents a straightforward and sustainable solution to the universal oracle problem. It is anchored in a decentralized, collective rating mechanism, making it organically adaptive to supply real-world data to smart contracts, thereby functioning as a dynamic oracle network.

The reputation system's intrinsic advantage stems from its foundational principles. It endows participants with a robust and enduring incentive to rate transparently and honestly. This distinct methodology yields unparalleled advantages. For consumers, it acts as a guiding beacon, steering them to vendors who meet their anticipated standards. For vendors, it stands as a defensive shield against unfounded negative feedback from malevolent competitors, enabling them to foster and uphold their credibility. As the system magnetically draws an escalating number of consumers and vendors, its network effect burgeons. This amplifying network effect continually reinforces the oracle. Moreover, its adaptability and universality negate the need for complex, case-specific designs. As it intuitively attracts diverse users and data providers, the reputation system is poised to rival other oracles like Chainlink.

Furthermore, with the potential integration of cryptographic tools such as Ring Signatures and Zero-Knowledge Proofs (ZKPs) to the reputation system, the oracle offers enhanced privacy adaptability:

\begin{enumerate}
    \item \textbf{Ring Signature Oracles:} Ring Signatures allow for digital signing by a group member, disclosing that a group member signed it but concealing the signer's identity. Within oracles, this protects data providers' identities. Specifically, as oracle providers feed data to smart contracts, using Ring Signatures ensures that the data's specific provider remains undisclosed, offering oracle participants a layer of privacy.
        
    \item \textbf{ZKP Oracles:} ZKPs let a prover authenticate a statement's veracity to a verifier without revealing further relevant information about the statement. In oracles, ZKPs protect the data contents submitted to smart contracts, permitting data verification without unveiling the data's detailed specifics. By integrating ZKPs, the system can effortlessly merge with existing blockchain frameworks, ensuring data privacy without compromising its accuracy or truth.
\end{enumerate}

The evolving nature of the reputation system, driven by agent participation and feedback, epitomizes a self-tuning mechanism. Incorporating a Decentralized Autonomous Organization (DAO) in future phases will endow the oracle with perpetual evolution, distinguishing it from static oracle systems reliant on manual interventions or enhancements.

\subsubsection{Advantageous Scenarios for the Universal Sustainable Oracle}
\begin{enumerate}
    \item \textbf{Privacy-Sensitive Applications:} Integration of Ring Signatures and ZKPs makes this system ideal for applications demanding high data confidentiality, such as banking transactions, confidential business dealings, legal documents, medical records, and private voting mechanisms.
    
    \item \textbf{Decentralized Finance (DeFi):} Precise data is the lifeblood of DeFi. The reputation-based oracle can enhance financial protocols with reliable market data, optimizing the efficacy of platforms related to lending, borrowing, trading, or even risk management and insurance.
    
    \item \textbf{Supply Chain Management:} The reputation system, with its emphasis on transparency and accuracy, can ensure product authenticity throughout supply chains. Industries prioritizing authenticity like agriculture, pharmaceuticals, electronics, luxury goods, and even art and antiques can greatly benefit.
\end{enumerate}

In summary, rooted in blockchain principles and supplemented with ZKPs, the everlasting reputation system offers a comprehensive solution to the pertinent oracle dilemma. As blockchain garners wider acceptance, the need for a trustworthy oracle will surge, positioning this model as a leading contender that harmoniously blends data privacy, reliability, and durability.

\section{Future Investigations} \label{sec:future_investigations}

%Additional and Multiple Ratings; Early Detection of Potential Problems; Adaptive Incentive Mechanisms for Post-Launch Stage; Perpetually Self-Adaptively-Evolving Universal Oracle

As our research journey navigates the intricate tapestry of the reputation system and its myriad facets, there remain certain areas poised for deeper exploration in subsequent phases of our investigation. Our anticipation is to cast our intellectual net even wider, to capture the more nuanced dynamics and emergent properties that our initial inquiry might have only touched upon. Firstly, we delve into the realm of \textbf{Additional and Multiple Ratings}, seeking to comprehend how multiple layers of feedback can enrich the analytical depth of our system. Proceeding from there, the lens of our inquiry shifts to an anticipatory stance, focusing on the \textbf{Early Detection of Potential Problems} — a proactive measure to ensure the seamless functioning of the reputation mechanism. As we transition into the practicalities of system deployment, the \textbf{Adaptive Incentive Mechanisms for Post-Launch Stage} promises a set of guiding principles and methodologies tailored for the evolving needs of a live system. Lastly, but by no means least, our exploration culminates in the conceptualization of a \textbf{Perpetually Self-Adaptively-Evolving Universal Oracle}, envisaging a system that not only evolves but also continually refines itself, embodying the principles of perpetual learning and adaptation.

\subsection{Additional and Multiple Ratings} \label{subsec:additional_ratings}
In our subsequent studies, we aim to delve deeper into the intricacies of a singular agent assigning multiple ratings to a counterpart. We are conceptualizing an innovative weighting formula that, rather than merely focusing on staking quantity, places added emphasis on recent ratings. Such a revision is aimed at capturing the fluidity of actions and potential shifts in an agent's viewpoint over time.

To operationalize this, we propose the incorporation of indices representing an agent's historical actions, coupled with a decay function to account for intervals between successive ratings. Our objective with this methodology is to temper the impact of aged ratings, acknowledging that the context of the scrutinized action may transform. To ensure its feasibility, we plan to employ MCSs within an expansive parameter spectrum. This will entail examining an array of decay functions related to diverse time spans.

\subsection{Early Detection of Potential Problems} \label{subsec:early_detection}

\subsubsection{Manual Intervention During Testnets} \label{subsubsec:Manual}
Despite the inherent design of the reputation system, which instinctively promotes agent integrity, and its satisfactory performance in simulations, it might be prudent to retain the option for manual interventions, particularly during the testnet phase. Potential indicators, like abrupt transaction surges or atypical staking by an agent, could be warning signs of looming issues or exploitation. Given the feedback dynamics of reputation systems, user reviews and ratings might act as preliminary alert signals. For example, a sharp decline in an agent's reputation could necessitate in-depth scrutiny. This can be optimally executed if the manual intervention guidelines are unambiguous, thereby ensuring seamless coexistence with the autonomous functioning of the reputation system.

\subsubsection{Machine Learning Integration} \label{subsubsec:MachineLearning}
Employing advanced analytics and machine learning algorithms can facilitate real-time transaction pattern monitoring. By amalgamating continuous oversight with adaptive algorithms proficient in learning from past data, the system can proactively pinpoint and address concerns, fortifying the trustworthiness of the blockchain and all its stakeholders. This anticipatory stance not only reduces potential threats but also nurtures user trust, bolstering durability and adaptability in a swiftly changing digital landscape.

\subsection{Adaptive Incentive Mechanisms for Post-Launch Stage} \label{subsec:adaptive_incentives}

\subsubsection{Incentive Mechanisms for Early Problem Finders} \label{subsubsec:early_problem_detection}
To counteract the "Keynesian beauty contest" effect, as delineated by Keynes in 1936 \cite{Keynes1936}, we contemplate the application of skewed functions to endorse rating diversity. If the collective outcomes of this "contest" exceed a predetermined threshold, the system might autonomously incentivize divergent actions, aiming for recalibration. Designing such mechanisms necessitates meticulous scheming, exhaustive simulation-based validation, and thorough field testing to ensure interventions remain well-measured and intentional.

\subsubsection{Strengthening Community Participation} \label{subsubsec:community_participation}
Besides prioritizing early problem identification, it's essential to reward the astute minority who identify these issues promptly. Echoing sentiments from earlier chapters, a renewed focus will be directed towards championing community engagement. Establishing an organic incentive framework can ensure sustained competitiveness of the reputation system. This strategy not only lauds individuals for their proactive contributions but also consolidates the efficacy and integrity of the overarching system.

\subsubsection{Exploring Decentralized Autonomy through DAO} \label{subsec:dao}

The exploration of Decentralized Autonomous Organizations (DAOs) presents an innovative blueprint for a self-regulating, independent system. By their design, DAOs, anchored in predefined rules, circumvent centralized control, potentially ensuring unbiased decision-making mechanisms. These organizations can be tailored to autonomously reward contributors, manage resources, and make judgements rooted in the community's collective wisdom. Incorporating DAOs into our system's architecture might further bolster incentivized participation, aligning individual rewards with collective benefits.

\subsection{Perpetually Self-Adaptively-Evolving Universal Oracle} \label{subsec:eternal_system}

An astutely crafted reputation system with intrinsic incentive mechanisms for integrity ratings, coupled with the principles of DAO, paves the way for a perpetually self-operating, self-correcting, and adaptively evolving Universal Oracle. This union merges the immutable trust mechanism of a DAO with the dynamic adaptability of our model, crafting a system poised for long-term resilience and relevance. Users, being both beneficiaries and decision-makers regarding the oracle's direction, ensure each update carries the legacy of existing participants while extending to new ones. This reinforces the network effect and establishes barriers against imitators. Consequently, such a system, by continuously recalibrating its parameters and practices in response to evolving circumstances, can potentially maintain its integrity and relevance indefinitely, surpassing any other competitors.

\vspace{-6pt}
 
\bigskip
 
\clearpage
\bibliographystyle{unsrt}

\end{document}